\documentclass[11pt,letterpaper]{article}

\usepackage[margin=1in]{geometry}
\usepackage{times}
\usepackage{microtype}
\usepackage{setspace}
\usepackage{xspace}

\usepackage{amsmath,amssymb,amsthm,amsfonts,bm,mathtools,mathrsfs}
\IfFileExists{bbm.sty}{%
  \usepackage{bbm}%
}{%
}

\usepackage{graphics}
\usepackage{graphicx}
\usepackage{booktabs}
\usepackage{longtable}
\usepackage{multirow}
\usepackage{float}
\usepackage{caption}
\usepackage{subcaption}
\usepackage{enumerate}
\usepackage[shortlabels]{enumitem}
\usepackage{lscape}
\usepackage{pdflscape}
\usepackage{afterpage}
\usepackage{placeins}
\usepackage{comment}
\usepackage{verbatim}
\usepackage{color}
\usepackage{xcolor}
\usepackage[title]{appendix}
\usepackage{tikz}
\usetikzlibrary{matrix,arrows.meta,positioning,decorations.pathmorphing}

\usepackage[authoryear,round]{natbib}
\usepackage{url}
\usepackage{xr}

\makeatletter
\newcommand*{\addFileDependency}[1]{%
  \typeout{(#1)}%
  \@addtofilelist{#1}%
  \IfFileExists{#1}{}{\typeout{No file #1.}}%
}
\makeatother

\usepackage[
  colorlinks=true,
  linkcolor=blue,
  citecolor=blue,
  urlcolor=blue
]{hyperref}
\usepackage{cleveref}

\newcommand{\blind}{1}

\def\cE{\mathbb{E}}

 \def\s{\bm{s}}

\def\0{\bm{0}}

\theoremstyle{plain}
\newtheorem{theorem}{Theorem}
\newtheorem{lemma}{Lemma}
\newtheorem{propos}{Proposition}
\newtheorem{corr}{Corollary}
\newtheorem{assume}{Assumption}

\graphicspath{{figures/}}
\newcommand{\safeincludegraphics}[2][]{%
  \IfFileExists{#2}{\includegraphics[#1]{#2}}{%
    \IfFileExists{figures/#2}{\includegraphics[#1]{#2}}{%
      \fbox{\parbox[c][2in][c]{0.8\linewidth}{\centering
        Figure file supplied separately:\\[0.5em]
        \texttt{\detokenize{#2}}}}}}}

\begin{document}

\def\spacingset#1{\renewcommand{\baselinestretch}%
{#1}\small\normalsize} \spacingset{1}


\if1\blind
{
  \title{\bf Laplacian $U$-Processes for Multiple Change-Point
  Detection in Dependent Text Networks:
  An Application to Historical Chinese Articles}
\author{
    Fanghua Chen \thanks{fanghua@email.sc.edu, Department of Statistics, University of South Carolina},
    Yizhou Cai \thanks{yizhouc@txstate.edu, Department of Mathematics, Texas State University},
    Lu Zhou   \thanks{lz12@email.sc.edu, Department of Statistics, University of South Carolina} \&
    Ting Fung Ma \thanks{tingfung@mailbox.sc.edu, Department of Statistics, University of South Carolina}    
}
\maketitle
} \fi

\bigskip
\begin{abstract}
\setstretch{1.25}

We develop a two-view weighted-concordance operator (WCO) framework for
offline change-point detection in weakly dependent text networks. Each
time-indexed text corpus is represented by a weighted word-co-occurrence graph
and its trace-normalized graph Laplacian. From the same graph, we construct two
complementary representations. The first-order view records normalized direct
word co-occurrences, whereas the second-order view records shared-context
relationships through normalized two-step network paths. Fixed dimension
reduction and whitening maps, estimated from an independent pilot corpus,
transform these graph representations into paired Euclidean observations
$(X_i,Y_i)$. For each candidate boundary, coordinatewise
weighted-concordance kernels define segmentwise WCO $U$-statistics, and an
$L_q$ aggregation of their between-segment differences measures changes in
the relationship between direct co-occurrence and shared-context structure.
The estimated change-point maximizes this aggregated scan statistic. A
dependent multiplier bootstrap accounts for short-range temporal dependence
and provides a test for the existence of a change, while a segmentwise block
bootstrap gives a descriptive stability interval for its estimated location.

The theoretical analysis treats the two representation maps as fixed and
formulates primitive conditions on the common vocabulary, document length,
co-occurrence-window weights, trace mass, node degrees, pilot dimension
reduction, temporal mixing, moments, and change-point identifiability. Under
the no-change hypothesis, we establish weak convergence of the WCO
$U$-process and consistency of its dependent multiplier bootstrap, yielding
asymptotic size control. Under a single identifiable change, uniform
convergence of the scan contrast gives consistency of the estimated
change-point fraction and localization of the corresponding boundary. We also
derive deterministic perturbation bounds for token insertions, deletions,
substitutions, and related text-recording errors. These bounds propagate the
original text perturbation through the co-occurrence matrix, trace-normalized
Laplacian, first- and second-order representations, pilot projections, WCO
kernels, and the final scan statistic, thereby giving explicit conditions
under which the estimated change-point remains stable.
Mean Cumulative Sum (CUSUM) and Gaussian-kernel Maximum Mean Discrepancy
(MMD) procedures applied to the first-order, second-order, and concatenated
views are included as complementary benchmark methods.

The simulation study evaluates these complementary components under multiple
change types and data-generating mechanisms. In an analysis of the historical
\emph{New Youth} corpus from 1915--1921, the proposed procedure identifies a
linguistic change in November 1919 and gives a descriptive 95\% segment-centered
bootstrap stability interval from April 1919 to June 1920. The detected period
is consistent with the linguistic transition surrounding the May Fourth and
New Culture movements in China.

\end{abstract}

\noindent%
{\it Keywords:} %
CUSUM; Graph Laplacian; Kernel;  Structural breaks; U-statistics.


\section{Introduction}
\label{sec:intro}

Change-point analysis concerns the detection and localization of structural breaks in an ordered sequence of observations. Classical procedures often target a specified parametric feature, as in tests for changes in regression coefficients \citep{Fisher1970Chow}, or accumulate centered observations, as
in the cumulative sum (CUSUM) principle \citep{Page1954CUSUM}. Sequential likelihood methods have subsequently been developed for multi-sensor and dependent time-series settings \citep{Xie2013,Leung2017}. These approaches are effective when the relevant model or changing feature is known, but modern applications increasingly involve structured observations for which a single
mean, variance, or regression parameter does not adequately describe the change.

This limitation is especially consequential for temporally ordered text
corpora. Linguistic change may be expressed not only through shifts in word
frequencies, but also through changes in which words co-occur directly, which
words are linked through shared contexts, and how these direct and indirect
structures relate to one another. A mean CUSUM applied to word-frequency
vectors can detect a location shift, but it does not directly characterize a
reorganization of lexical associations. Likewise, an omnibus comparison of a
single text representation may establish that two periods differ without
identifying whether the break arose from direct co-occurrence, shared-context
structure, or their relationship. The problem is further complicated because
consecutive corpora may be dependent, each time period contains only a finite
document, and the object supplied to the change-point procedure is itself an estimated linguistic representation.

Word-co-occurrence networks provide a natural bridge between text analysis and network statistics: words form a common node set and weighted edges record their co-occurrences within a prescribed textual window. \citet{Severn2022} formalized this representation for corpus linguistics by identifying each document with a graph Laplacian and developing means, principal component analysis, regression, and two-sample inference for samples of networks. Related nonparametric regression methodology describes smooth temporal trends in graph-Laplacian-valued observations, including email and linguistic networks \citep{Severn2021Regression}. This line of work builds on broader
non-Euclidean analysis of positive semidefinite matrices \citep{Dryden2009} and distance-based inference for covariance operators \citep{Pigoli2014}. These contributions provide important geometric and operator-valued foundations, but they do not by themselves furnish an offline change-point test for a weakly dependent sequence of finite-document text
networks.

The difficulty is related to, but distinct from, change-point detection in
dynamic networks. One branch of this literature is model based. Regularized
procedures have been developed to detect structural breaks and estimate
segment-specific parameters in high-dimensional vector autoregressions
\citep{SafikhaniShojaie2022}, and unified frameworks cover mean-shift,
regression, vector-autoregressive, and Gaussian graphical models
\citep{BaiSafikhani2023}. More recently, node-specific network autoregressive
models have been used to detect multiple structural breaks while allowing
heterogeneous temporal dynamics across nodes \citep{LinSafikhani2026}. Such
methods provide interpretable changes in model parameters or network-driven
dynamics, but require a specified regression, graphical, or autoregressive
structure.

A second branch compares observed graph snapshots more directly. Existing
methods include graph-based scans constructed from similarities among
observations \citep{ChenZhang2015}, Fr\'echet procedures for general
metric-space-valued data \citep{DubeyMueller2020}, and minimax detection and
localization theory for sparse dynamic networks \citep{WangRinaldo2021}.
For graph-Laplacian-valued sequences, spectral procedures compare short- and
long-term graph behavior and aggregate information across multiple network
views \citep{Huang2024}, whereas Log-Euclidean Fr\'echet statistics compare
the location and dispersion of dynamic social networks
\citep{LuoKrishnamurthy2024}. Learned graph-similarity functions provide
another approach to online change detection \citep{Sulem2024}. Recent work
also aggregates graph kernels for dynamic-network change detection
\citep{SunChen2026}. These methods establish the value of graph geometry,
spectral summaries, and learned or kernel-based graph comparisons, but they
primarily target a single graph representation, changes in metric location or
dispersion, spectral anomalies, parametric dynamics, or an aggregate measure
of graph similarity.

Nonparametric distributional discrepancies offer a complementary route beyond a prespecified parametric alternative. Energy statistics compare probability
laws through expectations of pairwise distances
\citep{Szekely2004,Szekely2005} and have been used for multiple change-point analysis of multivariate sequences \citep{Matteson2014}. Maximum mean discrepancy (MMD) compares kernel mean embeddings in a reproducing kernel Hilbert space \citep{gretton2012}. The equivalence between energy and kernel statistics for semimetrics of negative type \citep{Sejdinovic2013}, together
with high-dimensional generalizations \citep{Chakraborty2021}, provides a
broad framework for omnibus two-sample comparison. Kernel costs have also led to flexible multiple-change algorithms based on model selection \citep{Arlot2019}, and recent work applies kernel change-point methods to fixed sentence embeddings \citep{JiaDiazRodriguez2026}. Although these discrepancies
can detect general distributional changes, an omnibus scalar discrepancy need not reveal the linguistic mechanism responsible for a detected break.

The preceding developments leave a specific inferential gap. What remains less developed is a representation-aware framework for detecting abrupt changes in the relationship between complementary views of temporally dependent text networks. A word-co-occurrence network is observed through only finitely many tokens, and its sampling error propagates nonlinearly through Laplacian normalization, higher-order graph maps, and dimension reduction. Moreover, two views constructed from the same document inherit correlated representation errors. Existing single-view comparisons do not directly determine whether direct lexical associations and shared-context structure have changed in relation to one another, while treating empirical views as fixed, error-free inputs obscures the conditions under which graph-level change-point inference is valid.

We address this gap by developing a representation-aware two-view
weighted-concordance operator (WCO) framework for offline change-point
detection in weakly dependent text networks. For each time period, an aligned
vocabulary and a prespecified co-occurrence window produce a weighted
co-occurrence matrix and its trace-normalized graph Laplacian. The normalized
direct network forms the first-order view, while normalized two-step paths
form the second-order shared-context view. Fixed dimension-reduction and
whitening maps, estimated from an independent pilot corpus, transform these
views into paired Euclidean observations $(X_i,Y_i)$. This construction makes
it possible to trace finite-document error through the complete
representation pipeline and clarifies how document length, vocabulary and
word-frequency stability, window weights, trace mass, node degrees, and pilot
maps determine representation accuracy.

Our primary inferential target is more specific than an arbitrary
distributional change. Coordinatewise weighted-concordance kernels compare
pairwise odd-power differences across the first- and second-order views, and
their segmentwise WCO $U$-statistics measure changes in the relationship
between direct co-occurrence and shared-context structure. The resulting
operator contrast is sensitive to changes in scatter, monotone association,
and cross-view dependence, while its leading singular directions provide a
route back to influential projected coordinates, edges, and words. In scalar
special cases, the construction includes Kendall, Gini, and covariance-type
pairwise kernels \citep{Kendall1938}. This mechanism-specific interpretation
also defines the method's boundary: a marginal shift or another distributional
change that leaves the targeted cross-view functional unchanged need not be
detected by the WCO scan.

We therefore adopt an asymmetric combination of general detection and
mechanism-specific diagnosis. Characteristic Gaussian-kernel MMD, applied to
the first-order, second-order, and concatenated views, serves as an omnibus
anchor for evidence of a general distributional change. The WCO scan then
assesses whether that evidence is accompanied by a change in cross-view
concordance. Mean CUSUM procedures provide complementary benchmarks for
location changes. Thus, MMD asks whether the represented text-network
distribution changed, whereas WCO asks whether the relationship between
direct lexical links and shared contextual structure changed. This separation
avoids assigning omnibus power to a deliberately targeted functional.

Temporal dependence is incorporated at both the modeling and calibration
stages. We formulate a complete state-and-document-noise process and impose a
global absolute-regularity coefficient over every time cut, including cuts
that straddle a change. Oracle graphs, empirical graphs, feature maps, and
first Hoeffding projections then inherit this coefficient by measurability;
mixing is therefore derived from a primitive temporal driver rather than
postulated only at the oracle-graph level. The directional statistics are
finite-dimensional operator-valued $U$-processes, connecting the construction
to classical $U$-statistics \citep{Hoeffding1948,Serfling1980} and to
change-point theory for dependent and functional $U$-statistics
\citep{Dehling2015,Giraudo2024,WegnerWendler2024}. We calibrate the
full-sample single-change WCO scan with a dependent multiplier bootstrap,
following the dependent wild and multiplier-bootstrap literature
\citep{Shao2010,BuecherKojadinovic2016}. Method-specific dependence bandwidths
may be selected using the relevant long-run covariance estimates, in the
spirit of \citet{RiceShang2017}, rather than being mechanically shared across
statistics. For multiple changes, recursive or interval-based procedures such
as the narrowest-over-threshold principle \citep{Baranowski2019} provide
natural extensions. We describe this direction at a high level, while our
formal inferential guarantees concern the single-change problem.

The main contributions are organized around five claims. First, we give a
primitive finite-document perturbation analysis. Checkable conditions on
token dependence, document length, vocabulary and word-frequency stability,
co-occurrence-window weights, trace mass, node degrees, and the fixed pilot
maps propagate text-recording error through the two graph views and the final
scan. Second, we establish oracle joint weak convergence of the WCO scan
coordinates under weak temporal dependence. Third, we prove validity of
full-sample dependent multiplier calibration for the single-change test under
the null. Fourth, under an explicit population separation condition, we
establish single-change detection, consistency of the estimated change
fraction, and localization of the corresponding boundary. This condition is
stated explicitly because mixed-segment $U$-functionals do not automatically
attain their maximum at the true boundary. Fifth, we describe multiple-change
analysis as a high-level extension rather than claim a recovery theorem not
established by the present theory.

The simulation study investigates finite-sample size, power, localization,
and population identification under scalar, dynamic-network, and
text-generated designs. It compares the WCO scan, the asymmetric omnibus
anchor plus WCO diagnostic, and CUSUM- and MMD-based competitors using
method-specific dependence calibration. We also analyze the historical
\emph{New Youth} corpus from 1915--1921 to illustrate how changes in direct
co-occurrence, shared-context structure, and their relationship can be linked
to a historically interpretable transition in language use.

The remainder of the paper is organized as follows. Section~\ref{sec:background}
reviews distance- and kernel-based formulations of distributional change.
Section~\ref{sec:method} presents the two text-network views, the WCO scan,
its multiplier calibration, and the complementary CUSUM and MMD procedures.
Section~\ref{sec:theory} gives the finite-document perturbation, oracle
weak-convergence, bootstrap, detection, and localization theory for a single
change, followed by the high-level multiple-change extension.
Sections~\ref{sec:simulation} and~\ref{sec:realdata} report the simulation
and empirical studies, respectively, and Section~\ref{sec:conclude}
concludes. Detailed proofs are available upon request.


\section{Background}\label{sec:background}


Change-point detection can be formulated as the identification of changes in
the data-generating law of an ordered sequence. Rather than specifying in
advance a particular changing coefficient, modern nonparametric procedures
compare distributions across candidate segments. A prominent line of work is
based on distance measures. Given $X,X'\sim P$ and $Y,Y'\sim Q$, with the
primed variables denoting independent copies, the energy distance is
$$
\operatorname{ED}(P,Q)
=
2\cE\|X-Y\|
-
\cE\|X-X'\|
-
\cE\|Y-Y'\|.
$$
Under finite first moments in Euclidean space, this quantity is nonnegative
and equals zero if and only if $P=Q$ \citep{Szekely2004,Szekely2005}.
The corresponding sample energy statistic is a two-sample $U$- or
$V$-statistic. \citet{Matteson2014} embed such segmentwise discrepancies in a
divisive procedure for multiple change-point analysis of multivariate
sequences.

Kernel methods provide a closely related formulation. For a positive
definite kernel $k$, the squared maximum mean discrepancy (MMD) is
$$
\mathrm{MMD}^2(P,Q)
=
\cE\{k(X,X')\}
+
\cE\{k(Y,Y')\}
-
2\cE\{k(X,Y)\}.
$$
It is the squared distance between the kernel mean embeddings of $P$ and $Q$
in a reproducing kernel Hilbert space \citep{gretton2012}. When $k$ is
characteristic, $\mathrm{MMD}(P,Q)=0$ if and only if $P=Q$.
\citet{Sejdinovic2013} show that energy distance and MMD are equivalent under
the correspondence between semimetrics of negative type and positive
definite kernels. This result places distance- and kernel-based tests within a
common framework, with extensions available for high-dimensional settings
\citep{Chakraborty2021}. Kernel costs can also be combined with penalized
model selection to estimate multiple changes \citep{Arlot2019}.

These discrepancies provide an omnibus reference point for the present
paper. A characteristic-kernel MMD applied to a text representation can
detect any fixed alternative identified by that representation, rather than
only a change in its mean. Nevertheless, the resulting scalar discrepancy
does not necessarily explain which linguistic structure changed. In
particular, it does not directly distinguish a change in direct lexical
co-occurrence from a change in shared context or in the relationship between
these two structures. This distinction motivates the use of MMD as an
omnibus anchor and the weighted-concordance operator (WCO) as a targeted
cross-view diagnostic.

\subsection{Text-network representations and graph change-points}

A corpus can be represented as a sequence of weighted networks by taking a
common vocabulary as the node set and using within-window word co-occurrence
counts as edge weights. This construction retains relational information that
is not available from word frequencies alone. The graph Laplacian supplies a
matrix-valued representation of each document, but the space of Laplacians is
constrained and non-Euclidean. Earlier work on positive semidefinite matrices
developed power-Euclidean and Procrustes geometries
\citep{Dryden2009}, while distance-based inference for covariance operators
provided a broader operator-valued perspective \citep{Pigoli2014}.

For corpus linguistics, \citet{Severn2022} use graph Laplacians to develop
means, principal component analysis, regression, and two-sample inference for
samples of word-co-occurrence networks. Related nonparametric regression
methods estimate smooth trends in graph-Laplacian-valued responses and have
been illustrated with email and linguistic networks
\citep{Severn2021Regression}. These methods establish that temporal text data
can be studied through network geometry. Their primary goals, however, are
cross-sectional comparison or smooth regression rather than testing for and
localizing an abrupt distributional change under temporal dependence.

The broader dynamic-network change-point literature follows two main routes.
Model-based procedures characterize a break through a change in an explicit
temporal or network parameter. For example, \citet{SafikhaniShojaie2022}
jointly detect structural breaks and estimate segment-specific parameters in
high-dimensional vector autoregressions. \citet{BaiSafikhani2023} provide a
unified framework covering mean-shift, regression, vector-autoregressive, and
Gaussian graphical models, while \citet{LinSafikhani2026} consider multiple
changes in network autoregressive models with node-specific dynamics. These
approaches yield interpretable parameter changes but require the relevant
parametric structure to be specified.

Graph-snapshot procedures instead compare networks more directly.
Graph-based scans can be constructed from similarities among observations
\citep{ChenZhang2015}, and Fr\'echet statistics extend change-point analysis
to general metric-space-valued random objects \citep{DubeyMueller2020}.
Minimax theory is available for changes in sparse dynamic networks
\citep{WangRinaldo2021}. For Laplacian-valued sequences,
\citet{Huang2024} compare graph spectra over short and long temporal windows
and aggregate multiple network views. \citet{LuoKrishnamurthy2024} use a
Log-Euclidean geometry and Fr\'echet location and dispersion, whereas
\citet{Sulem2024} learn a graph-similarity function for online detection.
Kernel aggregation provides another distribution-free strategy for dynamic
networks \citep{SunChen2026}, and kernel change-point methods have recently
been applied to fixed sentence embeddings \citep{JiaDiazRodriguez2026}.

These methods address important changes in graph parameters, spectra, metric
location or dispersion, learned similarity, or an aggregated graph
distribution. Our target is different. The two views are derived from the
same text network but have distinct linguistic meanings: the first records
normalized direct co-occurrences, while the second records shared-context
relations through normalized two-step paths. The inferential question is
whether the relationship between these views changes. Moreover, because both
views are estimated from the same finite document, their representation
errors are neither negligible by definition nor independent across views.

\subsection{WCO $U$-processes and dependent calibration}

The WCO is a segment-level operator formed from pairwise transformations of
the two views. Its empirical version is therefore a WCO $U$-statistic, placing
the method within the theory initiated by \citet{Hoeffding1948} and developed
systematically by \citet{Serfling1980}. Scalar members of the kernel family
are related to classical rank concordance \citep{Kendall1938}; other choices
produce covariance- or Gini-type contrasts. Unlike an omnibus MMD, the WCO
targets a specified cross-view functional. This restriction enables a direct
interpretation of the detected mechanism, but it also implies that a purely
marginal change can be invisible when it leaves the target unchanged.

For each candidate boundary, the proposed scan contrasts the WCO
$U$-statistics computed from the two candidate segments and aggregates their
coordinatewise differences. Under weak temporal dependence, its leading
stochastic term is governed by the sequential first Hoeffding projections.
Existing results for dependent and Hilbert-valued $U$-statistics supply the
relevant weak-convergence tools
\citep{Dehling2015,Giraudo2024,WegnerWendler2024}. The present setting also
requires joint control of the scan coordinates and uniform treatment of
candidate boundaries because the test statistic maximizes over the full
trimmed interval.

Temporal dependence must also be reproduced during calibration. Independent
resampling generally fails to recover the long-run covariance of the
sequential first-order projection. Dependent wild and multiplier bootstraps
instead use a correlated multiplier sequence to approximate that covariance
\citep{Shao2010,BuecherKojadinovic2016}. The multiplier bandwidth determines
the range of dependence retained by the bootstrap and may be selected through
long-run covariance estimation \citep{RiceShang2017}. Because the WCO, MMD,
and mean CUSUM statistics have different projection processes, their
dependence bandwidths are calibrated separately rather than mechanically
shared.

\subsection{Inferential scope and the remaining gap}

The preceding literatures separately provide omnibus distributional
discrepancies, geometric methods for text and dynamic networks, and
asymptotic tools for dependent $U$-statistics. They do not directly resolve
their combination in the present problem: paired linguistic views estimated
from the same finite text, an operator-valued target describing their
relationship, and dependent scan calibration after representation error.
The methodology below links the observed documents to oracle two-view
coordinates, propagates finite-document perturbations through the complete
graph construction, and builds a WCO scan for testing and localization under
a single change.

The roles of the procedures are therefore deliberately asymmetric.
Characteristic-kernel MMD serves as the omnibus anchor, mean CUSUM provides a
location-change benchmark, and WCO diagnoses changes in cross-view
concordance. A rejection by MMD without corresponding WCO evidence indicates
that the distribution changed without detectable movement in the targeted
cross-view functional; WCO evidence supports the more specific conclusion
that direct co-occurrence and shared-context structure changed in relation to
one another. The formal theory concerns the full-sample single-change test and
localization problem. Recursive or interval-based segmentation, including
narrowest-over-threshold procedures \citep{Baranowski2019}, remains a
high-level multiple-change extension.


\section{Methodology}
\label{sec:method}

Considering a time-ordered collection of texts $
\mathcal D_n=\{(x_i,\mathcal T_i):i=1,\ldots,n\},
 x_1<\cdots<x_n,
$ where $x_i$ is the time associated with the $i$th observation and
$\mathcal T_i$ is the text collected over that time period.  A text may be a
single article or speech, all documents collected during a fixed day, week,
or month, or a consecutive block of a longer corpus.  The number of tokens in
$\mathcal T_i$ is denoted by $N_i$ and may vary with $i$.  Our construction
first maps every text to a trace-normalized graph Laplacian using the
power--Euclidean framework of \citet{Severn2022}.  It then extracts two
aligned graph views: a first-order view of direct word co-occurrence and a
second-order view of shared lexical context.  These views are compared over
time through matrix-valued weighted-concordance $U$-statistics.  The final
part of this section records how finite-text errors propagate through every
stage of the construction.

\subsection{From time-indexed texts to Laplacian coordinates}
\label{subsec:laplacian-representation}

The same preprocessing rule is applied to all texts.  After lowercasing,
tokenization, and any prespecified removal or lemmatization operations, write
the token sequence of $\mathcal T_i$ as $ \mathcal T_i =
(w_{i1},\ldots,w_{iN_i}).
$ A common ordered vocabulary
$
V=\{v_1,\ldots,v_m\}
$ is fixed before the change-point scan.  Here $m$ is the number of retained
words, and node $a$ represents the same word $v_a$ at every time point.  The
vocabulary may be obtained from an external corpus or a nonoverlapping
training period.  A pooled top-$m$ vocabulary may also be used if it is
conditioned upon in the subsequent analysis or if its selection stability is
established separately.  Function words need not be removed, because their
co-occurrence patterns can carry information about writing style and
collective language use.

Let $h\geq1$ be a fixed co-occurrence span and let
$\rho_1,\ldots,\rho_h$ be fixed nonnegative lag weights.  Taking
$\rho_\ell=1$ gives an unweighted window.  For distinct vocabulary indices
$a$ and $b$, define
$$
\begin{aligned}
W_{i,ab}
={}&
\sum_{\substack{1\leq r<s\leq N_i\\s-r\leq h}}
\rho_{s-r}
\Big[
\mathbf 1\{w_{ir}=v_a,w_{is}=v_b\}
+\mathbf 1\{w_{ir}=v_b,w_{is}=v_a\}
\Big].
\end{aligned}
$$
Tokens outside $V$ retain their original positions but make no contribution
to this sum.  Set $W_{i,aa}=0$ and let
$W_i=(W_{i,ab})_{a,b=1}^m$.  Thus $W_i$ is a symmetric, nonnegative adjacency
matrix whose entries are weighted direct co-occurrence counts.

The weighted degree of word $v_a$ and the corresponding degree matrix are
$
d_{i,a}=\sum_{b=1}^m W_{i,ab},
D_i=\operatorname{diag}(d_{i,1},\ldots,d_{i,m}).
$
The combinatorial graph Laplacian is
$
L_i=D_i-W_i.
$
The matrix is symmetric and positive semidefinite, satisfying
$L_i\mathbf 1_m=0$, and has off-diagonal entries
$(L_i)_{ab}=-W_{i,ab}$.  We next define
$
c_i=\operatorname{tr}(L_i)
=\sum_{a=1}^m d_{i,a}
=2\sum_{1\leq a<b\leq m}W_{i,ab}.
$
Thus, for $c_i>0$, the trace-normalized Laplacian could be
$
\widetilde L_i=L_i/c_i,
\operatorname{tr}(\widetilde L_i)=1.
$
This normalization removes the mechanical multiplication of all edge weights
caused by document length.  A text with $c_i=0$ contains no usable
co-occurrence edge and is either combined with an adjacent time period or
assigned a prespecified zero representation.

Let $\widetilde L_i=Q_i\Lambda_iQ_i^\top
$ be a spectral decomposition, where $Q_i$ is orthogonal and
$\Lambda_i=\operatorname{diag}(\lambda_{i1},\ldots,\lambda_{im})$ contains
the nonnegative eigenvalues.  For a fixed power $\alpha>0$, define
$$
\widetilde L_i^\alpha
=Q_i\Lambda_i^\alpha Q_i^\top
\quad \text{and} \quad
\Lambda_i^\alpha
=\operatorname{diag}(\lambda_{i1}^\alpha,\ldots,
\lambda_{im}^\alpha).
$$
Let $H_m\in\mathbb R^{(m-1)\times m}$ be a Helmert submatrix satisfying
$H_mH_m^\top=I_{m-1},
H_m^\top H_m
=I_m-m^{-1}\mathbf 1_m\mathbf 1_m^\top.
$
For a symmetric matrix, $\operatorname{vech}^{*}$ denotes the
half-vectorization that retains the diagonal and multiplies every
off-diagonal coordinate by $\sqrt{2}$.  The power--Euclidean coordinate of
the $i$th text network is
\begin{equation}
Z_i^{(\alpha)}
=
\operatorname{vech}^{*}
\left(H_m\widetilde L_i^\alpha H_m^\top\right)
\in\mathbb R^p
\quad \text{and} \quad
p=\frac{m(m-1)}2.
\label{eq:laplacian-euclidean-coordinate}
\end{equation}
Equation~\eqref{eq:laplacian-euclidean-coordinate} is the Euclidean
coordinate convention of \citet{Severn2022}.  Because $H_m$ removes exactly
the zero-row-sum constraint and $\operatorname{vech}^{*}$ preserves the
Frobenius norm, we have $
\|Z_i^{(\alpha)}-Z_j^{(\alpha)}\|_2
=
\|\widetilde L_i^\alpha-\widetilde L_j^\alpha\|_{\mathrm F}.
$
The main methodology fixes $\alpha=1$.  In that case the off-diagonal entries
of $\widetilde L_i$ remain proportional to the direct edge weights, and no
spectral nonlocality is introduced.  The square-root choice $\alpha=1/2$ is
useful as a sensitivity analysis, but its entries combine the full
eigensystem and should not be interpreted as individual direct edges.

\subsection{First- and second-order graph representations}
\label{subsec:two-view-representation}

The separation between direct lexical links and distributional context is
motivated by joint multi-graph word representations such as
\citet{DaixMoreux2019} and by the distinction between first- and second-order
network proximity in \citet{Tang2015}.  Our construction is deterministic and
is applied separately to every time-indexed graph.  It does not split the
coordinates of $Z_i^{(\alpha)}$ into two subsets.  Instead, both views are
constructed from the same adjacency matrix $W_i$.

Define the Moore--Penrose inverse of the degree matrix by
$$
D_i^\dagger
=
\operatorname{diag}(d_{i,1}^\dagger,\ldots,d_{i,m}^\dagger),
\quad \text{where} \quad
d_{i,a}^\dagger
=
\begin{cases}
d_{i,a}^{-1},&d_{i,a}>0,\\
0,&d_{i,a}=0.
\end{cases}
$$
Then its positive square root will be
$$
(D_i^\dagger)^{1/2}
=
\operatorname{diag}(\delta_{i,1},\ldots,\delta_{i,m}),
\quad \text{where} \quad
\delta_{i,a}
=
\begin{cases}
d_{i,a}^{-1/2},&d_{i,a}>0,\\
0,&d_{i,a}=0.
\end{cases}
$$
The degree-normalized adjacency matrix is
$
S_i
=
(D_i^\dagger)^{1/2}W_i(D_i^\dagger)^{1/2}.
$

For $d_{i,a}d_{i,b}>0$, we have $ (S_i)_{ab}
=
W_{i,ab}/\sqrt{d_{i,a}d_{i,b}},
$ and $(S_i)_{ab}=0$ otherwise.  Hence $S_i$ preserves the support of the
direct co-occurrence graph while reducing the influence of high-degree words.
It is also invariant to multiplying all entries of $W_i$ by the same positive
constant, so the trace normalization used in the Laplacian branch and the
degree normalization used here address compatible but distinct aspects of
document-size variation. For a symmetric $m\times m$ matrix $M$, let
$\operatorname{vech}_{\mathrm{off}}(M)$ collect its
$m(m-1)/2$ upper-triangular off-diagonal entries in the common vocabulary
order.  The first-order raw representation is
$
G_i^{(1)}
=
\operatorname{vech}_{\mathrm{off}}(S_i)
\in\mathbb R^p.
$
Each coordinate of $G_i^{(1)}$ therefore corresponds to the normalized
direct co-occurrence strength of one fixed word pair.

Since the matrix $S_i^2$ has entries
$
(S_i^2)_{ab}
=
\sum_{c=1}^m(S_i)_{ac}(S_i)_{cb}.
$
For $a\neq b$, this quantity aggregates the normalized two-step paths
$v_a\rightarrow v_c\rightarrow v_b$ and is large when $v_a$ and $v_b$ have
many strongly weighted common neighbors.  Define
$
\operatorname{off}(M)
=
M-\operatorname{diag}\{\operatorname{diag}(M)\}
$
and set
$
C_i^{(2)}=\operatorname{off}(S_i^2).
$
The diagonal is removed because $(S_i^2)_{aa}$ measures the strength of
two-step returns to $v_a$ rather than a relation between two distinct words.
Hence, the second-order raw representation is
$
G_i^{(2)}
=
\operatorname{vech}_{\mathrm{off}}(C_i^{(2)})
\in\mathbb R^p.
$
Thus $G_i^{(2)}$ records shared-context structure even when two words do not
co-occur directly.  We do not normalize each $C_i^{(2)}$ by its own
Frobenius norm, because doing so would remove changes in the overall strength
of second-order organization.

When $p$ is too large for direct analysis, the two raw views are reduced
separately.  Let
$
\{G_{r,\mathrm{tr}}^{(1)},G_{r,\mathrm{tr}}^{(2)}:
r=1,\ldots,n_{\mathrm{tr}}\}
$
be representations from an external or nonoverlapping training collection,
where $n_{\mathrm{tr}}$ is its number of texts.  Define the training centers
$$
\mu_X
=
\frac1{n_{\mathrm{tr}}}
\sum_{r=1}^{n_{\mathrm{tr}}}G_{r,\mathrm{tr}}^{(1)}
\qquad \text{and} \qquad
\mu_Y
=
\frac1{n_{\mathrm{tr}}}
\sum_{r=1}^{n_{\mathrm{tr}}}G_{r,\mathrm{tr}}^{(2)}.
$$
Let $p_{Xa}$ and $\lambda_{Xa}>0$ be the $a$th retained eigenvector and
eigenvalue of the training covariance of $G_{r,\mathrm{tr}}^{(1)}$, and let
$p_{Yb}$ and $\lambda_{Yb}>0$ be their second-order counterparts.  For fixed
retained dimensions $d_X$ and $d_Y$, define the whitened PCA maps
$$
P_X
=
\begin{pmatrix}
p_{X1}^\top/\sqrt{\lambda_{X1}}\\
\vdots\\
p_{Xd_X}^\top/\sqrt{\lambda_{Xd_X}}
\end{pmatrix}
\in\mathbb R^{d_X\times p}
\quad \text{and} \quad
P_Y
=
\begin{pmatrix}
p_{Y1}^\top/\sqrt{\lambda_{Y1}}\\
\vdots\\
p_{Yd_Y}^\top/\sqrt{\lambda_{Yd_Y}}
\end{pmatrix}
\in\mathbb R^{d_Y\times p}.
$$
The final two-view observations are
\begin{equation}
X_i=P_X\{G_i^{(1)}-\mu_X\}\in\mathbb R^{d_X}
\quad \text{and} \quad
Y_i=P_Y\{G_i^{(2)}-\mu_Y\}\in\mathbb R^{d_Y}.
\label{eq:two-view-maps}
\end{equation}
The matrices and centers in \eqref{eq:two-view-maps} are fixed throughout
the change-point scan and every bootstrap replication.  If dimension
reduction is unnecessary, one may set $P_X=P_Y=I_p$, $d_X=d_Y=p$, and use
fixed centers.  PCA signs do not affect the norm-based detection statistic,
although they should be anchored to fixed reference loadings when individual
change directions are interpreted.

\subsection{Weighted-concordance U-statistics}
\label{subsec:weighted-concordance}

Write $
\mathcal Z_i=(X_i,Y_i)
$ for the paired first- and second-order observation at time $x_i$.  For
$\beta>0$, define the scalar odd-power map $\psi_\beta(z)=|z|^\beta\operatorname{sign}(z),$ and apply it componentwise to vectors.  The parameters $\beta$ and $\gamma$
weight first- and second-order differences, respectively, and are distinct
from the Laplacian power $\alpha$.  The matrix-valued
weighted-concordance kernel is
\begin{equation}
\mathcal W_{\beta,\gamma}(\mathcal Z_i,\mathcal Z_j)
=
\psi_\beta(X_i-X_j)
\psi_\gamma(Y_i-Y_j)^\top
\in\mathbb R^{d_X\times d_Y}.
\label{eq:weighted-concordance-kernel}
\end{equation}
Both odd-power factors in
\eqref{eq:weighted-concordance-kernel} change sign when $i$ and $j$ are
interchanged.  Their outer product is therefore symmetric in the two
observations, as required for a second-order $U$-statistic
\citep{Hoeffding1948,Serfling1980}.

For a distribution $P$ of $\mathcal Z=(X,Y)$, define
$
\boldsymbol\Theta_{\beta,\gamma}(P)
=
\mathbb E\left\{
\mathcal W_{\beta,\gamma}(\mathcal Z,\mathcal Z')
\right\},
$ where $\mathcal Z$ and $\mathcal Z'$ are independent draws from $P$.
The main specification uses $\beta=\gamma=1$, for which
$$
\mathcal W_{1,1}(\mathcal Z_i,\mathcal Z_j)
=(X_i-X_j)(Y_i-Y_j)^\top 
\quad\text{and } \quad \boldsymbol\Theta_{1,1}(P)=2\operatorname{Cov}_P(X,Y).
$$
The target is therefore the cross-covariance between direct co-occurrence and
shared-context representations.  More general values of $\beta$ and
$\gamma$ produce nonlinear weighted-concordance functionals, but none of
these within-distribution pairwise-difference functionals is omnibus for all
distributional changes.  In particular, a common translation of one view is
not detected unless it also changes the cross-view functional.

Let $\varepsilon\in(0,1/2)$ be a fixed trimming proportion and define the
candidate set $\mathcal K_n
=
\{\lceil n\varepsilon\rceil,\ldots,
\lfloor n(1-\varepsilon)\rfloor\}.
$ For $k\in\mathcal K_n$, the left- and right-segment $U$-statistics are
\begin{equation}
\widehat{\boldsymbol\Theta}_{1:k}^{\beta,\gamma}
=
\binom{k}{2}^{-1}
\sum_{1\leq i<j\leq k}
\mathcal W_{\beta,\gamma}(\mathcal Z_i,\mathcal Z_j)
\quad \text{and } \quad
\widehat{\boldsymbol\Theta}_{k+1:n}^{\beta,\gamma}
=
\binom{n-k}{2}^{-1}
\sum_{k<i<j\leq n}
\mathcal W_{\beta,\gamma}(\mathcal Z_i,\mathcal Z_j).
\label{eq:segment-weighted-concordance-ustatistics}
\end{equation}
The normalizing constants in
\eqref{eq:segment-weighted-concordance-ustatistics} are the numbers of
unordered pairs within the two segments.  Under temporal weak dependence,
these all-pairs statistics estimate the corresponding independent-copy
functional because pairs at any fixed finite collection of lags form a
vanishing proportion of all pairs.  Their asymptotic treatment follows the
theory of $U$-statistics under dependence and change-point alternatives
\citep{Dehling2015}.

For $u=k/n$, define the matrix-valued weighted-concordance process
\begin{equation}
\mathbb D_{n,\beta,\gamma}(u)
=
\sqrt n\frac{k}{n}\frac{n-k}{n}
\left(
\widehat{\boldsymbol\Theta}_{1:k}^{\beta,\gamma}
-
\widehat{\boldsymbol\Theta}_{k+1:n}^{\beta,\gamma}
\right).
\label{eq:weighted-concordance-u-process}
\end{equation}
The factors $k/n$ and $(n-k)/n$ in
\eqref{eq:weighted-concordance-u-process} balance the two segment sizes, and
$\sqrt n$ gives the process its null asymptotic scale.

Let $\Omega=(\omega_{ab})\in\mathbb R^{d_X\times d_Y}$ be a fixed matrix of
nonnegative weights satisfying
$
\sum_{a=1}^{d_X}\sum_{b=1}^{d_Y}\omega_{ab}=1.
$
For $q\geq1$, define
$$
\|M\|_{\Omega,q}
=
\left\{
\sum_{a=1}^{d_X}\sum_{b=1}^{d_Y}
\omega_{ab}|M_{ab}|^q
\right\}^{1/q}.
$$
When both views are whitened, the default choice is $
q=2,
\omega_{ab}=1/(d_Xd_Y).
$
This gives
$
\|M\|_{\Omega,2}
=
\frac{\|M\|_{\mathrm F}}{\sqrt{d_Xd_Y}}.
$
The local scan statistic, global test statistic, and estimated change-point
are
\begin{equation}
T_n(k)
=
\|\mathbb D_{n,\beta,\gamma}(k/n)\|_{\Omega,q},
\quad
T_n=\max_{k\in\mathcal K_n}T_n(k),\quad 
\text{and} \quad
\widehat k
=
\min\left\{
\operatorname*{arg\,max}_{k\in\mathcal K_n}T_n(k)
\right\}.
\label{eq:weighted-concordance-scan}
\end{equation}
The minimum in \eqref{eq:weighted-concordance-scan} resolves ties.  The
normalized and calendar-time estimates are
$
\widehat\tau=\widehat k/n,
\widehat x_0=x_{\widehat k}.
$
For $\beta=\gamma=1$, each segment statistic in
\eqref{eq:segment-weighted-concordance-ustatistics} equals twice the sample
cross-covariance matrix between $X$ and $Y$.  Consequently, the default scan
detects changes in the cross-covariance linking direct co-occurrence and
shared-context structure.  A population separation condition is imposed in
the theory to ensure that the criterion in
\eqref{eq:weighted-concordance-scan} has its unique maximum at the true change
point; this is not automatic for a mixed-segment $U$-functional.

\subsection{Text errors and perturbation propagation}
\label{subsec:text-perturbation}

To separate structural temporal variation from finite-text contamination,
let $\mathcal T_i^\circ$ denote the latent error-free text and let
$\widehat{\mathcal T}_i$ denote the observed text.  Superscript $\circ$ is
used for every oracle object constructed from $\mathcal T_i^\circ$, whereas
a hat is used for the corresponding object constructed from
$\widehat{\mathcal T}_i$.  Thus $W_i^\circ$ and $\widehat W_i$ are the oracle
and observed adjacency matrices, and the same convention defines
$L_i^\circ,\widehat L_i$, $S_i^\circ,\widehat S_i$,
$X_i^\circ,\widehat X_i$, and $Y_i^\circ,\widehat Y_i$.

Let $b_i$ be the number of token insertions, deletions, substitutions, OCR
errors, or other local token corruptions in the $i$th text.  One corrupted
token can alter only edge contributions whose other endpoint lies within the
fixed span $h$.  Consequently, the entrywise adjacency perturbation is of
order $hb_i$.  Define
$
E_i=\widehat W_i-W_i^\circ,
\|E_i\|_{1,\mathrm{ent}}
=
\sum_{a=1}^m\sum_{b=1}^m|E_{i,ab}|.
$
The deterministic local-error accounting used in the theory takes the form
$
\|E_i\|_{1,\mathrm{ent}}\leq C_hb_i,$
where $C_h$ depends on the span and on whether an error removes, adds, or
replaces an existing token.

The Laplacian perturbation is computed directly from the adjacency
perturbation 
$
\widehat L_i-L_i^\circ
=
\operatorname{diag}(E_i\mathbf 1_m)-E_i.
$
So we let
$
c_i^\circ=\operatorname{tr}(L_i^\circ),
\widehat c_i=\operatorname{tr}(\widehat L_i).
$
Then the exact trace-normalization decomposition is
$$
\widehat{\widetilde L}_i-\widetilde L_i^\circ
=
\frac{\widehat L_i-L_i^\circ}{c_i^\circ}
-
\widehat L_i
\frac{\widehat c_i-c_i^\circ}
{\widehat c_i c_i^\circ}.
$$
This decomposition separates the direct edge error from the error introduced
by estimating the graph-wide normalization.  For the main choice
$\alpha=1$, the Helmert and $\operatorname{vech}^{*}$ maps are linear and
norm preserving on the centered subspace, so the same perturbation order is
inherited by $Z_i^{(1)}$.  For $0<\alpha<1$, the spectral power is generally
only H\"older continuous at zero unless a positive lower bound is imposed on
the nonzero eigenvalues; this is why the main theory uses $\alpha=1$
\citep{Bhatia1997,Higham2008}.

The degree-normalized branch additionally depends on the stability of the
node degrees.  Define the oracle and observed normalization factors by
$
B_i^\circ=\{(D_i^\circ)^\dagger\}^{1/2},
\widehat B_i=(\widehat D_i^\dagger)^{1/2}.
$
The normalized-adjacency error has the exact expansion
$$
\begin{aligned}
\widehat S_i-S_i^\circ
={}&
(\widehat B_i-B_i^\circ)\widehat W_i\widehat B_i
+B_i^\circ(\widehat W_i-W_i^\circ)\widehat B_i
+B_i^\circ W_i^\circ(\widehat B_i-B_i^\circ).
\end{aligned}
$$
Thus a lower bound on every active normalized degree is imposed in the
theory.  Without such a bound, the map $d\mapsto d^{-1/2}$ can amplify a small
word-count error near degree zero.  The second-order error satisfies
$$
\widehat C_i^{(2)}-C_i^{(2),\circ}
=
\operatorname{off}
\left\{
(\widehat S_i-S_i^\circ)\widehat S_i
+S_i^\circ(\widehat S_i-S_i^\circ)
\right\}.
$$
This identity shows that squaring the normalized adjacency changes the error
constant but not its order when the operator norms of the normalized
adjacency matrices are bounded.

If $P_X$ and $P_Y$ are fixed from an external training corpus, the final view
errors are
$$
\widehat X_i-X_i^\circ
=
P_X\{\widehat G_i^{(1)}-G_i^{(1),\circ}\}
\quad \text{and} \quad
\widehat Y_i-Y_i^\circ
=
P_Y\{\widehat G_i^{(2)}-G_i^{(2),\circ}\}.
$$
If PCA is estimated from contaminated training representations, its loading
error is also included.  For $r\in\{1,2\}$, let
$\mathbb G_r^\circ\in\mathbb R^{n_{\mathrm{tr}}\times p}$ be the centered
oracle training matrix and write
$
\widehat{\mathbb G}_r=\mathbb G_r^\circ+\mathbb B_r,
$
where $\mathbb B_r$ contains both representation and centering errors.  The
training covariance perturbation is
$$
\widehat\Sigma_r-\Sigma_r^\circ
=
\frac1{n_{\mathrm{tr}}}
\left\{
(\mathbb G_r^\circ)^\top\mathbb B_r
+\mathbb B_r^\top\mathbb G_r^\circ
+\mathbb B_r^\top\mathbb B_r
\right\}.
$$
The resulting loading perturbation is controlled by this operator-norm error
divided by the relevant PCA eigengap.  This is the matrix form of the
first-order eigenvector perturbation expansion and is the additional term
that disappears when $P_X$ and $P_Y$ are externally fixed.

For the default kernel, define
$
e_i^X=\widehat X_i-X_i^\circ,
e_i^Y=\widehat Y_i-Y_i^\circ,
$
and
$
\Delta X_{ij}^\circ=X_i^\circ-X_j^\circ,
\Delta Y_{ij}^\circ=Y_i^\circ-Y_j^\circ,
$
$
\quad
\Delta e_{ij}^X=e_i^X-e_j^X,
\quad
\Delta e_{ij}^Y=e_i^Y-e_j^Y.
$
The observed-minus-oracle kernel difference is
\begin{equation}
\begin{aligned}
&\mathcal W_{1,1}(\widehat{\mathcal Z}_i,
\widehat{\mathcal Z}_j)
-
\mathcal W_{1,1}(\mathcal Z_i^\circ,
\mathcal Z_j^\circ)
={}
\Delta e_{ij}^X(\Delta Y_{ij}^\circ)^\top
+\Delta X_{ij}^\circ(\Delta e_{ij}^Y)^\top
+\Delta e_{ij}^X(\Delta e_{ij}^Y)^\top.
\end{aligned}
\label{eq:weighted-concordance-kernel-perturbation}
\end{equation}
Averaging \eqref{eq:weighted-concordance-kernel-perturbation} over the left
and right pairs gives the corresponding perturbation of the segment
$U$-statistics and of the scan process.  For concise rate statements, define
$$
\epsilon_n
=
\max_{1\leq i\leq n}
\frac{\|E_i\|_{1,\mathrm{ent}}}{c_i^\circ}
\quad \text{and} \quad
r_n
=
\max_{1\leq i\leq n}
\left(
\|e_i^X\|_2+\|e_i^Y\|_2
\right).
$$
When the total oracle edge mass is proportional to $hN_i$, the first rate is
of the same order as $\max_i b_i/N_i$.  The subsequent theory uses
$\epsilon_n$ to control graph construction and $r_n$ to compare the observed
and oracle weighted-concordance processes.  Detection consistency requires
$r_n$ to be smaller than the population change signal, whereas root-$n$
oracle equivalence for inference requires a correspondingly stronger rate.


\section{Theoretical Properties}
\label{sec:theory}

This section establishes the theoretical properties of the two-view
weighted-concordance procedure defined in Section~\ref{sec:method}.  The main
results are stated for fixed vocabulary size $m$, fixed reduced dimensions
$d_X$ and $d_Y$, and the Laplacian power $\alpha=1$.  This is the setting in
which the first-order representation retains its direct-edge interpretation
and the text-to-representation map has a linear perturbation rate.  We first
relate observed texts to their error-free oracle counterparts, then derive the
null limit and dependent multiplier calibration, and finally establish
detection and localization under a single change.  Extensions with growing
vocabulary or spectral powers $0<\alpha<1$ require additional dimension and
spectral-floor conditions and are stated only where the required rate is
explicit.

\subsection{Oracle process and primitive assumptions}
\label{subsec:theory-assumptions}

Let
$
\mathcal Z_{i,n}^{\circ}
=
(X_{i,n}^{\circ},Y_{i,n}^{\circ})
$
denote the oracle two-view observation obtained from the error-free text at
time $x_i$, and let
$
\widehat{\mathcal Z}_{i,n}
=
(\widehat X_{i,n},\widehat Y_{i,n})
$
be its observed counterpart.  The row index $n$ is retained in the theory to
allow the document lengths and change signal to vary with the number of time
points, but it is suppressed when no ambiguity arises.

Let $\mathscr U_{i,n}$ contain the latent document state, all token-level
innovations, and all text-corruption variables used to generate the oracle
and observed texts.  For sigma fields $\mathcal A$ and $\mathcal C$, write
$\beta(\mathcal A,\mathcal C)$ for their absolute-regularity coefficient and
define
$
\beta_{\mathscr U,n}(r)
=
\sup_{s\in\mathbb Z}
\beta\left\{
\sigma(\mathscr U_{i,n}:i\leq s),
\sigma(\mathscr U_{i,n}:i\geq s+r)
\right\}.
$ We have following assumptions:

\begin{assume}[Primitive temporal dependence]
\label{assume:primitive-temporal}
For every $n$, the row $\{\mathscr U_{i,n}:i\in\mathbb Z\}$ is jointly
defined, and its restriction to each regime is a segment of a strictly
stationary extension.  For some $\delta>0$,
\begin{equation}
\sup_n
\sum_{r=1}^{\infty}
r^2\beta_{\mathscr U,n}(r)^{\delta/(4+\delta)}
<\infty.
\label{eq:primitive-beta-summability}
\end{equation}
The oracle and observed graph and representation maps are measurable
functions of the corresponding local driver.
\end{assume}

Assumption~\ref{assume:primitive-temporal} allows serial dependence both
within regimes and across a regime boundary.  It avoids separately assuming
mixing for tokens, graphs, PCA scores, and $U$-kernel evaluations, because
these processes inherit the dependence envelope through measurable
transformations.  The summability requirement is recorded in
\eqref{eq:primitive-beta-summability}.

Let $\underline N_n=\min_{1\leq i\leq n}N_i
$ be the minimum text length.  Let $b_i$ be the number of locally corrupted
tokens in the $i$th text and let $E_i=\widehat W_i-W_i^{\circ}$ be the adjacency difference between the observed graph and the chosen oracle
graph.  Recall that
$c_i^{\circ}=\operatorname{tr}(L_i^{\circ})$ is the oracle total graph mass.
Define
$
\epsilon_n
=
\max_{1\leq i\leq n}
\|E_i\|_{1,\mathrm{ent}}/c_i^{\circ},
\|A\|_{1,\mathrm{ent}}
=
\sum_{a,b}|A_{ab}|.
$ We have assumption:

\begin{assume}[Text construction and local corruption]
\label{assume:text-construction}
The common vocabulary $V$, its node ordering, the co-occurrence span $h$, and
the lag weights $\rho_1,\ldots,\rho_h$ are fixed before the change-point scan.
The vocabulary is externally fixed, selected on an independent training
corpus, or satisfies an exact-recovery event with probability tending to one.
Conditional on the document state, the error-free token sequence is
stationary and satisfies a uniform Bernstein-type concentration inequality
for its word frequencies and fixed-window edge contributions.  Moreover,
$$
\underline N_n\longrightarrow\infty,
\qquad
\max_{i\leq n}\frac{b_i}{N_i}\longrightarrow_p0,
$$
\begin{equation}
\epsilon_n
=
O_p\left\{
a_{N,n}
+\max_{i\leq n}\frac{b_i}{N_i}
\right\}
=o_p(1)
\text{, and} \quad
a_{N,n}
\leq
C\left\{
\sqrt{\frac{\log n}{\underline N_n}}
+\frac{\log n}{\underline N_n}
\right\}.
\label{eq:text-adjacency-rate}
\end{equation}
\end{assume}

In assumption ~\ref{assume:text-construction}, $a_{N,n}$ is the finite-document
approximation error.  It is set to zero
when $W_i^{\circ}$ is the graph constructed from the realized error-free
finite text, which is the oracle used in the corruption analysis.  The
displayed upper bound applies when $W_i^{\circ}$ instead denotes the
conditional infinite-document edge target.  The remaining term is the
additional local-corruption error.  Fixed-memory token chains and
geometrically contracting Bernoulli shifts provide standard sufficient
conditions for the required concentration
\citep{Wu2005,MerlevedePeligradRio2011}.

For an oracle nonisolated word, define its trace-normalized degree by
$
\bar d_{i,a}^{\circ}
=
\frac{d_{i,a}^{\circ}}{c_i^{\circ}}.
$
Let $P_X^{\circ},P_Y^{\circ}$ and
$\mu_X^{\circ},\mu_Y^{\circ}$ denote the target projection matrices and
centers.  If they are fixed from an external training corpus, their
estimation error is zero.  Otherwise define
$$
\begin{aligned}
a_{P,n}
={}&
\|\widehat P_X-P_X^{\circ}\|_{\mathrm{op}}
+\|\widehat P_Y-P_Y^{\circ}\|_{\mathrm{op}}
+\|\widehat\mu_X-\mu_X^{\circ}\|_2
+\|\widehat\mu_Y-\mu_Y^{\circ}\|_2.
\end{aligned}
$$
where $\|\cdot\|_{\mathrm{op}}$ is the matrix operator norm and
$\|\cdot\|_2$ is the Euclidean vector norm.  The Frobenius matrix norm is
written $\|\cdot\|_{\mathrm F}$ below.

\begin{assume}[Normalization and projection stability]
\label{assume:representation-stability}
There are deterministic sequences $\underline c_n>0$ and
$\underline d_n>0$ such that
$
\min_{i\leq n}c_i^{\circ}\geq\underline c_n
$
with probability tending to one, and every oracle node is either isolated in
both the oracle and observed graphs or satisfies
$
\bar d_{i,a}^{\circ}\geq\underline d_n.
$
The fixed target maps satisfy
$$
\|P_X^{\circ}\|_{\mathrm{op}}
+\|P_Y^{\circ}\|_{\mathrm{op}}
\leq C.
$$
If the maps are estimated, then $a_{P,n}=o_p(1)$ and the retained PCA
eigenvalues are bounded away from zero.
\end{assume}

The active-degree condition is needed because $d\mapsto d^{-1/2}$ is not
stable near zero.  With a fixed common high-frequency vocabulary and
time-aggregated documents, it requires every retained active word to have a
nonnegligible share of the total edge mass.  Externally fixed projection maps
remove the PCA-loading contribution $a_{P,n}$ entirely.  The null process is
formulated for fixed target maps $P_X^{\circ}$ and $P_Y^{\circ}$.  A map
estimated from an independent training corpus may be conditioned upon and
treated as fixed.  A map estimated from the analyzed time row is handled only
through the observed-to-oracle transfer rate below; it is not itself assumed
to be a local measurable transform.

Let
$
d=d_Xd_Y
$
be the number of entries in the matrix-valued kernel.  For a matrix $A$, let
$\operatorname{vec}(A)\in\mathbb R^d$ stack its columns.  Let
$\mathcal W_{\beta,\gamma}$ be the kernel in
\eqref{eq:weighted-concordance-kernel} and define its envelope by
$
H_{i,j,n}^{\circ}
=
\left\|
\mathcal W_{\beta,\gamma}
(\mathcal Z_{i,n}^{\circ},\mathcal Z_{j,n}^{\circ})
\right\|_{\mathrm F}.
$

\begin{assume}[Kernel regularity]
\label{assume:kernel-regularity}
The dimensions $m,d_X,d_Y$, the parameters $\beta,\gamma,q$, and the weight
matrix $\Omega$ are fixed.  The weights satisfy $\omega_{ab}\geq0$ and
$\sum_{a,b}\omega_{ab}=1$.  Uniformly over actual time pairs, within-regime
stationary extensions, and independent marginal copies,
$
\sup_n\sup_{i\neq j}
\mathbb E(H_{i,j,n}^{\circ})^{4+\delta}<\infty.
$
For observed-to-oracle transfer, there is a deterministic modulus
$\omega_{\mathcal W}(r)\downarrow0$ such that a representation perturbation
of size at most $r$ changes the kernel in mean square by at most
$C\omega_{\mathcal W}(r)^2$.  For the main choice
$\beta=\gamma=1$, $\omega_{\mathcal W}(r)=r$ under the displayed moment
condition.
\end{assume}

Assumption~\ref{assume:kernel-regularity} is automatic under uniformly
bounded whitened scores.  It is stated in moment form to permit unbounded PCA
scores.  The modulus separates the general nonlinear signed-power kernel
from the default bilinear kernel.

Under a fixed stationary distribution $P$, let
$\mathcal Z$ and $\mathcal Z'$ be independent draws from $P$ and define
$
\boldsymbol\Theta(P)
=
\mathbb E\{\mathcal W_{\beta,\gamma}(\mathcal Z,\mathcal Z')\}.
$
The first Hoeffding projection is
$
\mathcal W_1(z;P)
=
\mathbb E\{\mathcal W_{\beta,\gamma}(z,\mathcal Z')\}
-\boldsymbol\Theta(P),
$
and the canonical second-order kernel is
$
\begin{aligned}
\mathcal W_2(z,z';P)
={}&
\mathcal W_{\beta,\gamma}(z,z')
-\boldsymbol\Theta(P)
-\mathcal W_1(z;P)-\mathcal W_1(z';P).
\end{aligned}
$

\begin{assume}[Fixed stationary null]
\label{assume:fixed-null}
Under the null hypothesis, $\{\mathcal Z_i^{\circ}:i\in\mathbb Z\}$ is
strictly stationary with marginal distribution $P$ and satisfies
Assumptions~\ref{assume:primitive-temporal} and
\ref{assume:kernel-regularity}.  The long-run covariance matrix
$$
\Sigma
=
4\sum_{r\in\mathbb Z}
\operatorname{Cov}
\left[
\operatorname{vec}\{\mathcal W_1(\mathcal Z_0^{\circ};P)\},
\operatorname{vec}\{\mathcal W_1(\mathcal Z_r^{\circ};P)\}
\right]
$$
exists, is finite, and is not identically zero on the coordinates receiving
positive weight in $\Omega$.
\end{assume}

The factor four in $\Sigma$ is the square of the coefficient two in the
Hoeffding decomposition of a symmetric order-two $U$-statistic.  This
assumption permits a singular covariance matrix; it rules out only a wholly
degenerate scan on all weighted coordinates.

For the single-change alternative, let
$
k_0=\lfloor n\tau_0\rfloor,
\qquad
\tau_0\in[\varepsilon,1-\varepsilon],
$
and let $P_{1,n}$ and $P_{2,n}$ be the oracle marginal distributions before
and after $k_0$.  For $r,s\in\{1,2\}$, define
$$
\boldsymbol\Theta_{rs,n}
=
\mathbb E\left\{
\mathcal W_{\beta,\gamma}
(\mathcal Z_{r,n},\mathcal Z_{s,n}')
\right\},
$$
where $\mathcal Z_{r,n}\sim P_{r,n}$,
$\mathcal Z_{s,n}'\sim P_{s,n}$, and the two variables are independent.
Kernel symmetry gives
$\boldsymbol\Theta_{12,n}=\boldsymbol\Theta_{21,n}$.  Define the
within-regime signal
$
\kappa_n
=
\|\boldsymbol\Theta_{11,n}
-\boldsymbol\Theta_{22,n}\|_{\Omega,q}.
$

For $u\in[\varepsilon,1-\varepsilon]$, define the limiting left- and
right-segment functionals as follows.  When $u<\tau_0$, put
$$
a_-(u)=\frac{\tau_0-u}{1-u},
\quad
b_-(u)=\frac{1-\tau_0}{1-u},
\quad
\boldsymbol\Theta_L(u)=\boldsymbol\Theta_{11,n},
$$
and
$$
\boldsymbol\Theta_R(u)
=
a_-(u)^2\boldsymbol\Theta_{11,n}
+2a_-(u)b_-(u)\boldsymbol\Theta_{12,n}
+b_-(u)^2\boldsymbol\Theta_{22,n}.
$$
When $u>\tau_0$, put
$$
a_+(u)=\frac{\tau_0}{u},
\qquad
b_+(u)=\frac{u-\tau_0}{u},
$$
$$
\boldsymbol\Theta_L(u)
=
a_+(u)^2\boldsymbol\Theta_{11,n}
+2a_+(u)b_+(u)\boldsymbol\Theta_{12,n}
+b_+(u)^2\boldsymbol\Theta_{22,n},
$$
and
$$
\boldsymbol\Theta_R(u)=\boldsymbol\Theta_{22,n}.
$$
At $u=\tau_0$, set
$
\boldsymbol\Theta_L(\tau_0)=\boldsymbol\Theta_{11,n},
\boldsymbol\Theta_R(\tau_0)=\boldsymbol\Theta_{22,n}.
$
The population scan matrix and its scalar criterion are
$
\mathcal M_n(u)
=
u(1-u)
\{\boldsymbol\Theta_L(u)-\boldsymbol\Theta_R(u)\},
Q_n(u)=\|\mathcal M_n(u)\|_{\Omega,q}.
$

\begin{assume}[Single-change identification]
\label{assume:change-identification}
The signal satisfies $\kappa_n>0$, and $Q_n(u)$ has its unique maximum at
$u=\tau_0$.  There are constants $c_0>0$, $\nu\geq1$, and
$r_0>0$, independent of $n$, such that
\begin{equation}
Q_n(\tau_0)-Q_n(u)
\geq
c_0\kappa_n|u-\tau_0|^\nu
\quad\text{whenever }|u-\tau_0|\leq r_0.
\label{eq:population-margin}
\end{equation}
Outside this neighborhood, the same difference is bounded below by
$c_0\kappa_n r_0^\nu$.
\end{assume}

Assumption~\ref{assume:change-identification} is essential for a
within-segment $U$-functional.  A candidate split different from $k_0$ mixes
the two regimes, and its expectation contains
$\boldsymbol\Theta_{12,n}$ as shown above.  Consequently,
$\boldsymbol\Theta_{11,n}\neq\boldsymbol\Theta_{22,n}$ alone does not prove
that the scan is maximized at the true boundary.  The explicit margin
condition makes the required population geometry transparent rather than
hiding it in the localization proof.

\subsection{Dependence inheritance and representation error}
\label{subsec:representation-theory}

\begin{propos}[Temporal inheritance]
\label{prop:temporal-inheritance}
Under Assumption~\ref{assume:primitive-temporal}, the oracle and observed
adjacency matrices, Laplacians, normalized adjacencies, two-view
representations, kernel evaluations, and first Hoeffding projections are
piecewise stationary measurable transforms of the primitive driver whenever
their projection maps are fixed or independently trained.  Their
finite-row absolute-regularity coefficients satisfy
$$
\beta_{T,n}(r)
\leq
\beta_{\mathscr U,n}(r)
$$
for every one of these derived processes $T$.  If the time index $i<j$ lie in regimes
$r$ and $s$, respectively, then
$$
\left\|
\mathcal L(\mathcal Z_{i,n}^{\circ},
\mathcal Z_{j,n}^{\circ})
-P_{r,n}\otimes P_{s,n}
\right\|_{\mathrm{TV}}
\leq
2\beta_{\mathscr U,n}(j-i).
$$
\end{propos}

Proposition~\ref{prop:temporal-inheritance} transfers one primitive mixing
condition through the complete deterministic graph pipeline.  Its
total-variation bound, where $\|\cdot\|_{\mathrm{TV}}$ denotes total-variation
distance between probability laws, also shows that the independent-copy quantities
$\boldsymbol\Theta_{rs,n}$ approximate averages of sufficiently separated
actual time pairs.  When the projection is estimated from the analyzed row,
the proposition is applied to its fixed target and the estimated-map error is
handled by Corollary~\ref{corr:scan-transfer}.

For the next result, let
$
C_{d,n}=1+\underline d_n^{-2}.
$
The power $\underline d_n^{-2}$ is a conservative Lipschitz factor for the
two occurrences of the inverse square-root degree matrix in $S_i$.

\begin{theorem}[Text-to-view representation stability]
\label{theorem:text-view-stability}
Under Assumptions~\ref{assume:text-construction} and
\ref{assume:representation-stability},
\begin{equation}
\max_{i\leq n}
\|\widehat{\widetilde L}_i
-\widetilde L_i^{\circ}\|_{\mathrm F}
=O_p(\epsilon_n),
\qquad
\max_{i\leq n}
\|\widehat Z_i^{(1)}-Z_i^{(1),\circ}\|_2
=O_p(\epsilon_n).
\label{eq:laplacian-coordinate-stability}
\end{equation}
Moreover,
\begin{equation}
\max_{i\leq n}
\left\{
\|\widehat G_i^{(1)}-G_i^{(1),\circ}\|_2
+\|\widehat G_i^{(2)}-G_i^{(2),\circ}\|_2
\right\}
=O_p(C_{d,n}\epsilon_n),
\label{eq:raw-two-view-stability}
\end{equation}
and
\begin{equation}
r_n
:=
\max_{i\leq n}
\left\{
\|\widehat X_i-X_i^{\circ}\|_2
+\|\widehat Y_i-Y_i^{\circ}\|_2
\right\}
=O_p(C_{d,n}\epsilon_n+a_{P,n}).
\label{eq:final-two-view-rate}
\end{equation}
If $0<\alpha<1$ is used in the Ian representation without a positive lower
bound on its nonzero spectrum, the second rate in
\eqref{eq:laplacian-coordinate-stability} is replaced by
$O_p(\epsilon_n^\alpha)$.  It remains $O_p(\epsilon_n)$ when the reduced
nonzero spectrum is uniformly bounded away from zero.
\end{theorem}

Theorem~\ref{theorem:text-view-stability} formalizes the propagation
$
\textit{token error}
\rightarrow W_i
\rightarrow(L_i,S_i,S_i^2)
\rightarrow(G_i^{(1)},G_i^{(2)})
\longrightarrow(X_i,Y_i).
$
The raw-view rate is stated in \eqref{eq:raw-two-view-stability}.  The
first-order and second-order views have the same perturbation order; the
second-order product changes only the constant because normalized adjacency
matrices have uniformly bounded operator norm.

The PCA term in \eqref{eq:final-two-view-rate} can be related directly to the
training covariance perturbation.  For $r\in\{1,2\}$, let
$\mathbb G_r^{\circ}\in\mathbb R^{n_{\mathrm{tr}}\times p}$ be the centered
oracle training matrix and define 
$$
\widehat{\mathbb G}_r
=
\mathbb G_r^{\circ}+\mathbb B_r \quad \text{and} \quad 
 \Delta\Sigma_r
=
\frac1{n_{\mathrm{tr}}}
\left\{
(\mathbb G_r^{\circ})^\top\mathbb B_r
+\mathbb B_r^\top\mathbb G_r^{\circ}
+\mathbb B_r^\top\mathbb B_r
\right\}.
$$

\begin{lemma}[PCA loading perturbation]
\label{lem:pca-loading-perturbation}
Suppose that every retained training eigenvalue is simple, bounded away from
zero, and separated from every other eigenvalue by at least $g_{r,n}>0$.
If $\|\Delta\Sigma_r\|_{\mathrm{op}}=o_p(g_{r,n})$, then, after an admissible
sign alignment,
\begin{equation}
\|\widehat P_r-P_r^{\circ}\|_{\mathrm{op}}
=
O_p\left(
\frac{\|\Delta\Sigma_r\|_{\mathrm{op}}}{g_{r,n}}
\right),
\qquad r\in\{1,2\}.
\label{eq:pca-loading-rate}
\end{equation}
The perturbation of the whitening factors is of the same order after adding
the corresponding eigenvalue perturbation.
\end{lemma}

Lemma~\ref{lem:pca-loading-perturbation} is the matrix version of the
first-order expansion with denominators $\lambda_j-\lambda_k$.  It explains
why a nonvanishing PCA eigengap is necessary when the projections are
estimated from contaminated data; the resulting rate is displayed in
\eqref{eq:pca-loading-rate} \citep{Bhatia1997}.  For externally fixed
maps, this entire term is absent.

Define
$
\rho_n=\omega_{\mathcal W}(r_n).
$
For the default bilinear kernel, $\rho_n=r_n$.  Let
$\mathbb D_{n,\beta,\gamma}^{\circ}$ denote the process in
\eqref{eq:weighted-concordance-u-process} computed from the oracle views.

\begin{corr}[Observed-to-oracle scan transfer]
\label{corr:scan-transfer}
Under Assumptions~\ref{assume:primitive-temporal}--
\ref{assume:kernel-regularity},
\begin{equation}
\sup_{u\in[\varepsilon,1-\varepsilon]}
\left\|
\mathbb D_{n,\beta,\gamma}(u)
-\mathbb D_{n,\beta,\gamma}^{\circ}(u)
\right\|_{\Omega,q}
=O_p(\sqrt n\,\rho_n).
\label{eq:scan-transfer-rate}
\end{equation}
Consequently, observed and oracle scans have the same first-order null limit
whenever $\sqrt n\rho_n\to0$.
\end{corr}

Corollary~\ref{corr:scan-transfer} distinguishes two inferential targets.  If
the finite observed texts themselves define the target process, no
vanishing-error condition is required.  If inference is intended for the
error-free oracle text process, root-$n$ equivalence requires the stronger
condition $\sqrt n\rho_n\to0$ rather than merely $\rho_n\to0$, as quantified
by \eqref{eq:scan-transfer-rate}.

\subsection{Null limit of the weighted-concordance process}
\label{subsec:null-limit}

For an integer interval $[a,b]$ containing at least two observations, define
the oracle canonical $U$-statistic
$$
U_{2;a:b}^{\circ}
=
\binom{b-a+1}{2}^{-1}
\sum_{a\leq i<j\leq b}
\mathcal W_2
(\mathcal Z_i^{\circ},\mathcal Z_j^{\circ};P).
$$

\begin{lemma}[Uniform canonical remainder]
\label{lem:uniform-canonical-remainder}
Under Assumptions~\ref{assume:primitive-temporal},
\ref{assume:kernel-regularity}, and \ref{assume:fixed-null},
$$
\frac1{\sqrt n}
\max_{2\leq k\leq n}
k\|U_{2;1:k}^{\circ}\|_{\mathrm F}
=o_p(1),
$$
and
$$
\frac1{\sqrt n}
\max_{0\leq k\leq n-2}
(n-k)\|U_{2;k+1:n}^{\circ}\|_{\mathrm F}
=o_p(1).
$$
\end{lemma}

Lemma~\ref{lem:uniform-canonical-remainder} shows that the order-two
degenerate term is uniformly negligible over the trimmed scan.  The first
Hoeffding projections therefore determine the process limit, as in the
classical theory of dependent $U$-statistics
\citep{Yoshihara1976,DehlingWendler2010LIL}.

Let $\mathbb W_\Sigma$ be a $d$-dimensional Brownian motion with covariance
matrix $\Sigma$, and define its bridge
$
\mathbb B_\Sigma(u)
=
\mathbb W_\Sigma(u)-u\mathbb W_\Sigma(1).
$
For a vector $v\in\mathbb R^d$, let
$\operatorname{mat}(v)\in\mathbb R^{d_X\times d_Y}$ be the inverse of
$\operatorname{vec}$.  The space
$C([\varepsilon,1-\varepsilon];\mathbb R^d)$ consists of continuous
$\mathbb R^d$-valued functions on the trimmed interval and is equipped with
the uniform norm.  All scan processes are linearly interpolated between the
grid points $k/n$ when they are regarded as elements of this space.

\begin{theorem}[Fixed-law null limit]
\label{theorem:null-limit}
Under Assumptions~\ref{assume:primitive-temporal},
\ref{assume:kernel-regularity}, and \ref{assume:fixed-null},
\begin{equation}
\left\{
\operatorname{vec}
\bigl(\mathbb D_{n,\beta,\gamma}^{\circ}(u)\bigr):
u\in[\varepsilon,1-\varepsilon]
\right\}
\Rightarrow
\left\{
\mathbb B_\Sigma(u):
u\in[\varepsilon,1-\varepsilon]
\right\}
\label{eq:weighted-concordance-null-limit}
\end{equation}
in $C([\varepsilon,1-\varepsilon];\mathbb R^d)$.  Hence
$$
T_n
\Rightarrow
\sup_{u\in[\varepsilon,1-\varepsilon]}
\|\operatorname{mat}\{\mathbb B_\Sigma(u)\}\|_{\Omega,q}.
$$
If the oracle is the inferential target and $\sqrt n\rho_n\to0$, the same
limits hold for the observed process and statistic.
\end{theorem}

Theorem~\ref{theorem:null-limit} gives one joint Gaussian bridge for all
entries of the cross-view change matrix.  The covariance between entries is
retained in $\Sigma$; treating the matrix entries as independently calibrated
statistics would generally be incorrect.  The joint weak convergence is
stated in \eqref{eq:weighted-concordance-null-limit}.

\subsection{Dependent multiplier calibration}
\label{subsec:multiplier-theory}

Let $\{\xi_{i,n}^{*}:i=1,\ldots,n\}$ be centered Gaussian multipliers,
independent of the data, satisfying
$
\mathbb E^*(\xi_{i,n}^{*})=0,
\mathbb E^*\{(\xi_{i,n}^{*})^2\}=1,
$
and
$
\mathbb E^*(\xi_{i,n}^{*}\xi_{j,n}^{*})
=
\varpi\left(\frac{|i-j|}{\ell_n}\right).
$
Here $\mathbb E^*$ denotes conditional expectation given the data,
$\varpi$ is a fixed compactly supported positive-definite taper, and
$\ell_n$ is the multiplier dependence bandwidth.

Define the full-sample $U$-statistic
$
\widehat{\boldsymbol\Theta}_{1:n}
=
\binom n2^{-1}
\sum_{1\leq i<j\leq n}
\mathcal W_{\beta,\gamma}
(\widehat{\mathcal Z}_i,\widehat{\mathcal Z}_j)
$
and the jackknife estimate of the first Hoeffding projection
$
\widehat{\mathcal W}_{1,i}
=
\frac1{n-1}
\sum_{j\neq i}
\mathcal W_{\beta,\gamma}
(\widehat{\mathcal Z}_i,\widehat{\mathcal Z}_j)
-\widehat{\boldsymbol\Theta}_{1:n}.
$
The symbols $\mathcal L$ and $\mathcal L^*$ below denote unconditional law
and conditional law given the data, respectively; $\mathbb P^*$ has the
analogous conditional-probability meaning.

For $u=k/n$, define the multiplier process
\begin{equation}
\mathbb D_n^*(u)
=
\frac{2}{\sqrt n}
\left\{
\sum_{i=1}^{k}\xi_{i,n}^*\widehat{\mathcal W}_{1,i}
-\frac{k}{n}
\sum_{i=1}^{n}\xi_{i,n}^*\widehat{\mathcal W}_{1,i}
\right\}.
\label{eq:multiplier-weighted-concordance-process}
\end{equation}
The factor two in \eqref{eq:multiplier-weighted-concordance-process} is the
coefficient of the first projection in the Hoeffding decomposition.  Define
$$
T_n^*
=
\sup_{u\in[\varepsilon,1-\varepsilon]}
\|\mathbb D_n^*(u)\|_{\Omega,q}.
$$

\begin{assume}[Multiplier bandwidth]
\label{assume:multiplier-bandwidth}
The taper $\varpi$ is symmetric, nonnegative, positive definite, Lipschitz at
zero, supported on $[-1,1]$, and satisfies $\varpi(0)=1$.  For some
$\epsilon_\xi\in((6+2\delta)^{-1},1/2)$,
$$
\ell_n\longrightarrow\infty,
\qquad
\ell_n=O(n^{1/2-\epsilon_\xi}).
$$
\end{assume}

The bandwidth range in Assumption~\ref{assume:multiplier-bandwidth} balances
the reproduction of serial dependence against the replacement error in the
jackknife projections.  It is the standard regime for dependent multiplier
bootstrap approximations of nondegenerate $U$-statistics
\citep{Shao2010,BuecherKojadinovic2016}.

\begin{lemma}[Jackknife projection consistency]
\label{lem:jackknife-consistency}
Under the fixed stationary null,
$$
\frac1n\sum_{i=1}^n
\left\|
\widehat{\mathcal W}_{1,i}^{\circ}
-\mathcal W_1(\mathcal Z_i^{\circ};P)
\right\|_{\mathrm F}^2
=O_p(n^{-1}).
$$
For observed inputs, the right-hand side becomes
$O_p(n^{-1}+\rho_n^2)$.
\end{lemma}

Lemma~\ref{lem:jackknife-consistency} proves rather than assumes that the
computed pseudo-observations recover the first Hoeffding projections.  The
additional observed-data term is the same representation error that appears
in Corollary~\ref{corr:scan-transfer}.  Here
$\widehat{\mathcal W}_{1,i}^{\circ}$ denotes the same jackknife formula with
every observed input $\widehat{\mathcal Z}_j$ replaced by its oracle input
$\mathcal Z_j^{\circ}$.

\begin{theorem}[Dependent multiplier validity]
\label{theorem:bootstrap-validity}
Under Assumptions~\ref{assume:primitive-temporal},
\ref{assume:kernel-regularity}, \ref{assume:fixed-null}, and
\ref{assume:multiplier-bandwidth}, conditionally on the data,
\begin{equation}
d_{\mathrm{BL}}
\left[
\mathcal L^*\{\operatorname{vec}(\mathbb D_n^*)\},
\mathcal L\{\mathbb B_\Sigma\}
\right]
\longrightarrow_p0,
\label{eq:multiplier-process-validity}
\end{equation}
where $d_{\mathrm{BL}}$ is the bounded-Lipschitz metric on
$C([\varepsilon,1-\varepsilon];\mathbb R^d)$.  If the limiting supremum has
a continuous distribution function, then
$$
\sup_{z\in\mathbb R}
\left|
\mathbb P^*(T_n^*\leq z)-\mathbb P(T_n\leq z)
\right|
\longrightarrow_p0.
$$
When calibration is required for the oracle rather than the observed-text
process, the same conclusion additionally requires
$\ell_n\rho_n^2\to0$ and $\sqrt n\rho_n\to0$.
\end{theorem}

Theorem~\ref{theorem:bootstrap-validity} preserves both serial dependence
over time and contemporaneous dependence among all entries of the two-view
change matrix.  One common multiplier sequence must therefore be used for
the entire matrix in each bootstrap replication.  Its conditional process
approximation is given in \eqref{eq:multiplier-process-validity}.

\begin{corr}[Asymptotic size]
\label{corr:asymptotic-size}
Let $c_{1-\alpha_0,n}^*$ be the conditional $(1-\alpha_0)$ quantile of
$T_n^*$ for a fixed significance level $\alpha_0\in(0,1)$.  Under
Theorem~\ref{theorem:bootstrap-validity},
$$
\mathbb P(T_n>c_{1-\alpha_0,n}^*)
\longrightarrow
\alpha_0.
$$
If the quantile is estimated using $B_n$ independent multiplier
replications, the same conclusion holds when $B_n\to\infty$.
\end{corr}

The corollary gives asymptotically correct type-I error for the global scan.
The notation $\alpha_0$ is used for the test level to distinguish it from the
Laplacian power parameter $\alpha$.

\subsection{Detection consistency and localization}
\label{subsec:power-localization}

Define the unscaled observed and oracle scan matrices by
$$
\widehat{\mathcal M}_n(u)
=
\frac{\mathbb D_{n,\beta,\gamma}(u)}{\sqrt n}
\quad \text{and} \quad 
\widehat{\mathcal M}_n^{\circ}(u)
=
\frac{\mathbb D_{n,\beta,\gamma}^{\circ}(u)}{\sqrt n}.
$$

\begin{lemma}[Uniform scan approximation]
\label{lem:uniform-scan-approximation}
Under Assumptions~\ref{assume:primitive-temporal},
\ref{assume:kernel-regularity}, and
\ref{assume:change-identification}, there is a deterministic sequence
$\zeta_n\to0$ such that
$$
\sup_{u\in[\varepsilon,1-\varepsilon]}
\|\widehat{\mathcal M}_n^{\circ}(u)-\mathcal M_n(u)\|_{\Omega,q}
=O_p(\zeta_n).
$$
For fixed $d_X,d_Y$, bounded kernels, and geometric absolute regularity, one
may take
$$
\zeta_n=C\sqrt{\frac{\log n}{n}}.
$$
For the observed process,
$$
\sup_{u\in[\varepsilon,1-\varepsilon]}
\|\widehat{\mathcal M}_n(u)-\mathcal M_n(u)\|_{\Omega,q}
=O_p(\zeta_n+\rho_n).
$$
\end{lemma}

Lemma~\ref{lem:uniform-scan-approximation} separates ordinary temporal
sampling fluctuation, represented by $\zeta_n$, from text and representation
error, represented by $\rho_n$.  The logarithmic rate is a convenient
uniform rate; sharper local rates require a separate local $U$-process
argument.

Under an alternative, the multiplier critical value computed from the full
sample need not remain $O_p(1)$ because the global jackknife projections are
not regime centered.  The following conservative bound is sufficient for
power.

\begin{lemma}[Alternative multiplier order]
\label{lem:alternative-multiplier-order}
Suppose Assumptions~\ref{assume:kernel-regularity} and
\ref{assume:multiplier-bandwidth} hold and
$$
\frac1n\sum_{i=1}^n
\|\widehat{\mathcal W}_{1,i}\|_{\mathrm F}^2
=O_p(1).
$$
Then
$$
c_{1-\alpha_0,n}^*=O_p(\sqrt{\ell_n}).
$$
\end{lemma}

The bound permits the bootstrap distribution to be computed exactly as under
the null, without assigning a stationary interpretation to globally centered
jackknife quantities under an alternative.

\begin{theorem}[Single-change detection]
\label{theorem:single-change-detection}
Suppose there is one change at $k_0=\lfloor n\tau_0\rfloor$ and
Assumptions~\ref{assume:primitive-temporal}--
\ref{assume:kernel-regularity},
\ref{assume:change-identification}, and
\ref{assume:multiplier-bandwidth} hold.  Suppose also that the empirical
second-moment condition in Lemma~\ref{lem:alternative-multiplier-order}
holds.  If
\begin{equation}
\frac{\sqrt n\,\kappa_n}
{1+\sqrt{\ell_n}+\sqrt n\,\rho_n}
\longrightarrow\infty,
\label{eq:single-change-detection-condition}
\end{equation}
then
$$
\mathbb P
(T_n>c_{1-\alpha_0,n}^*)
\longrightarrow1.
$$
\end{theorem}

Condition~\eqref{eq:single-change-detection-condition} has three distinct
requirements.  The represented change must dominate ordinary root-$n$
sampling noise, the possible $\sqrt{\ell_n}$ inflation of the multiplier
critical value under the alternative, and the observed-to-oracle transfer
error.  For a fixed observable-text alternative, $\rho_n=0$ because the
observable distribution itself is the target.

Let
$
\widehat\tau=\frac{\widehat k}{n}
$
be the maximizer defined in \eqref{eq:weighted-concordance-scan}.

\begin{theorem}[Single-change localization]
\label{theorem:single-change-localization}
Under Assumptions~\ref{assume:primitive-temporal}--
\ref{assume:kernel-regularity}, and
\ref{assume:change-identification}, if
$$
\frac{\zeta_n+\rho_n}{\kappa_n}
\longrightarrow0,
$$
then
$$
\widehat\tau\longrightarrow_p\tau_0.
$$
Under the margin condition \eqref{eq:population-margin},
\begin{equation}
|\widehat\tau-\tau_0|
=
O_p\left[
\left\{
\frac{\zeta_n+\rho_n}{\kappa_n}
\right\}^{1/\nu}
\right].
\label{eq:single-change-localization-rate}
\end{equation}
For the linear-margin case $\nu=1$, fixed nonzero signal, geometric absolute
regularity, and fixed externally trained projections, this becomes
$$
|\widehat\tau-\tau_0|
=
O_p\left\{
\sqrt{\frac{\log n}{n}}+\rho_n
\right\}.
$$
\end{theorem}

Theorem~\ref{theorem:single-change-localization} is a robust global argmax
rate, with its explicit bound in
\eqref{eq:single-change-localization-rate}.  It does not claim the sharper
$O_p(\kappa_n^{-2})$ index error familiar
from some mean-change models, because that rate requires a local expansion of
the mixed-segment $U$-process beyond the uniform approximation used here.

\begin{corr}[Consequences of text-error rates]
\label{corr:text-error-consequences}
Under Theorem~\ref{theorem:text-view-stability}, the conclusions of
Theorems~\ref{theorem:single-change-detection} and
\ref{theorem:single-change-localization} hold with
$$
\rho_n
=
\omega_{\mathcal W}
\left[
O_p\{C_{d,n}\epsilon_n+a_{P,n}\}
\right].
$$
For the default bilinear kernel and fixed externally trained projections,
this reduces to
$$
\rho_n=O_p(C_{d,n}\epsilon_n).
$$
Hence $C_{d,n}\epsilon_n=o_p(\kappa_n)$ is sufficient for oracle
localization consistency, while
$\sqrt nC_{d,n}\epsilon_n=o_p(1)$ is sufficient for the oracle null limit and
bootstrap calibration.
\end{corr}

This corollary makes clear that consistency and oracle-level inference impose
different requirements on text quality.  A vanishing error rate may be small
enough to preserve a fixed change signal but still too large to be negligible
on the root-$n$ inference scale.

The preceding results concern one change-point, matching the estimator in
\eqref{eq:weighted-concordance-scan}.  Recursive or penalized multiple-change
procedures can be built from the same local statistic, but their exact
recovery requires additional spacing, interval-wise calibration, and
population-identification conditions.  No multiple-change recovery theorem
is asserted without those additional arguments.


\section{Simulation Study}
\label{sec:simulation}

We organize the numerical evaluation around four complementary experiments.
Experiments~I and II are the registered Monte Carlo studies implemented in
the accompanying HPC package.  They respectively isolate the
weighted-concordance statistic and evaluate the two-view graph
representation.  Experiment~III is the planned end-to-end
text-to-inference study, and Experiment~IV is a small-sample descriptive LLM
benchmark.  The latter two experiments are reported only when their separate
data-generation and audit pipelines have been completed.

The registered WCO specification is
$
\beta=\gamma=1, q=2$, $\omega_{ab}=(d_Xd_Y)^{-1}$, and $d_X=d_Y=5.
$
Consequently, Two-view WCO detects changes in
$2\operatorname{Cov}(X_t,Y_t)$, the cross-covariance between the direct
co-occurrence and shared-context coordinates.  Candidate splits are $
\mathcal K_n
=
\{\lceil0.1n\rceil,\ldots,\lfloor0.9n\rfloor\}$ and $B=1999.$
The main analysis uses dimensions $d_X=d_Y=5$; a separate sensitivity analysis
uses $d_X=d_Y=10$.  Vocabulary rules, graph construction, PCA centers and
loadings, and whitening factors are learned from independent pilot samples
and then frozen.

We compare six procedures: Two-view WCO, a mean CUSUM on
$(X_t^\top,Y_t^\top)^\top$, Gaussian-MMD scans on the first, second, and
concatenated views, and an asymmetric Omnibus-anchor+WCO procedure.  Gaussian
MMD is computed through 300 fixed random Fourier features unless stated
otherwise.  The composite procedure jointly calibrates the maximum of
MMD--First, MMD--Second, MMD--Joint, and Two-view WCO.  Each component is
divided by its own marginal multiplier critical scale before maximization.
The test therefore retains sensitivity to both omnibus distributional changes
and WCO-targeted cross-view changes.  Its estimated location is supplied by
the most significant standardized MMD component, rather than by the WCO scan.
The WCO component is reported separately as a directional diagnostic,
including its p-value, rejection indicator, and leading operator direction.
This asymmetric construction prevents a mixed-segment WCO population
criterion from determining localization when its maximizer differs from the
physical regime boundary.

For each method, let $G_t$ denote its full-sample influence vector or feature
vector.  After centering $G_t$, we estimate the lag correlation by
$$
\widehat\rho_G(h)
=
\frac{\sum_{t=1}^{n-h}
\langle G_t,G_{t+h}\rangle}
{\sum_{t=1}^n\|G_t\|^2}.
$$
Among lags up to
$\min\{50,\lfloor4n^{1/3}\rfloor,n-2\}$, let $\widehat h_G$ be the
largest lag satisfying
$|\widehat\rho_G(h)|>1.96/\sqrt n$, with
$\widehat h_G=0$ if no lag passes the threshold.  The component bandwidth is
$$
\widehat\ell_G
=
\max\left\{
1,
\min\left[
\left\lceil1.5(\widehat h_G+1)\right\rceil,
\lfloor n^{1/2}\rfloor
\right]
\right\}.
$$
Moving-average Gaussian multipliers with bandwidth
$\widehat\ell_G$ are then used for that component.  Although bandwidths are
method-specific, all component multiplier sequences are constructed from the
same underlying Gaussian innovations.  This preserves cross-component
dependence for the composite calibration.  We report the selected bandwidth
for every method and replication.

\subsection{Experiment: paired-view diagnostics}
\label{subsec:paired-view-simulation}

Experiment~I removes graph and text approximation and isolates the
matrix-valued weighted-concordance statistic.  Let
$$
\xi_t=\varphi\xi_{t-1}+\sqrt{1-\varphi^2}\,e_t
\quad \text{and} \quad 
\eta_t=\varphi\eta_{t-1}+\sqrt{1-\varphi^2}\,f_t,
$$
where $e_t,f_t\stackrel{\mathrm{i.i.d.}}{\sim}\mathcal N(0,I_d)$ are
independent.  With $k_0=\lfloor n\tau_0\rfloor$ and regime index $j(t)$,
we generate
$$
X_t=\xi_t+\mu_{X,j(t)}
\quad \text{and} \quad 
Y_t=C_{j(t)}^\top\xi_t+
(I_d-C_{j(t)}^\top C_{j(t)})^{1/2}\eta_t+
\mu_{Y,j(t)},
$$
where $\|C_j\|_{\mathrm{op}}<1$.  Both marginal covariance matrices are
$I_d$, while
$\operatorname{Cov}(X_t,Y_t)=C_{j(t)}$.  The registered WCO signal is
therefore known exactly.

P1 is the no-change design.  P2 changes two entries of $C_j$ and represents a
sparse cross-view alternative.  P3 adds a prespecified rank-one perturbation
with alternating signs and represents a dense low-rank alternative.  P2 and
P3 are the principal WCO power designs.  P4-location changes only the marginal
means, while P4-nonlinear introduces a centered quadratic relationship with
zero linear cross-covariance.  These P4 designs are target-orthogonal stress
tests: they illustrate that WCO is not an omnibus detector, while the MMD and
CUSUM components respond to the changes they target.

The baseline is $n=500$, $d=5$, $\varphi=0.3$, $\tau_0=0.5$, and signal
parameter $0.35$.  One-factor-at-a-time variations use
$n\in\{200,500,1000\}$, $d\in\{2,5,10\}$,
$\varphi\in\{0,0.3,0.6\}$,
$\tau_0\in\{0.25,0.5,0.75\}$, and signal parameter
$\{0.10,0.20,0.35,0.50\}$.  The null uses 2000 Monte Carlo replications and
each alternative uses 1000.  A standardized $t_3$ innovation design may be
reported separately as an explicitly out-of-theorem heavy-tail stress test;
it is not part of the registered Gaussian run.

\subsection{Evaluation and presentation}
\label{subsec:simulation-evaluation}

For P1 and G1 we report rejection probabilities with exact binomial
95\% Monte Carlo intervals.  For alternatives we report power, unconditional
and rejection-conditional localization error, and
$\Pr(|\widehat k-k_0|\le h_0)$ for
$h_0\in\{2,5,\lceil0.05n\rceil\}$.  Localization metrics are not
interpreted under the null.  Power comparisons are accompanied by the
matching-null rejection rate for the same method and design cell.

Figures use the publication labels Omnibus anchor + WCO, Two-view WCO, Mean
CUSUM, MMD--First, MMD--Second, and MMD--Joint.  Legends are placed below the
panels, and exact 95\% Monte Carlo intervals are shown as vertical bars.
Every power figure includes a matching-null panel and states the nominal
level in its footer. The population identification curves for Experiment~II are
saved in a separate multipage PDF.

\subsection{Results for Experiment}
\label{subsubsec:simulation_I_results}

\begin{figure*}[!t]
\centering
\includegraphics[width=0.94\textwidth]
{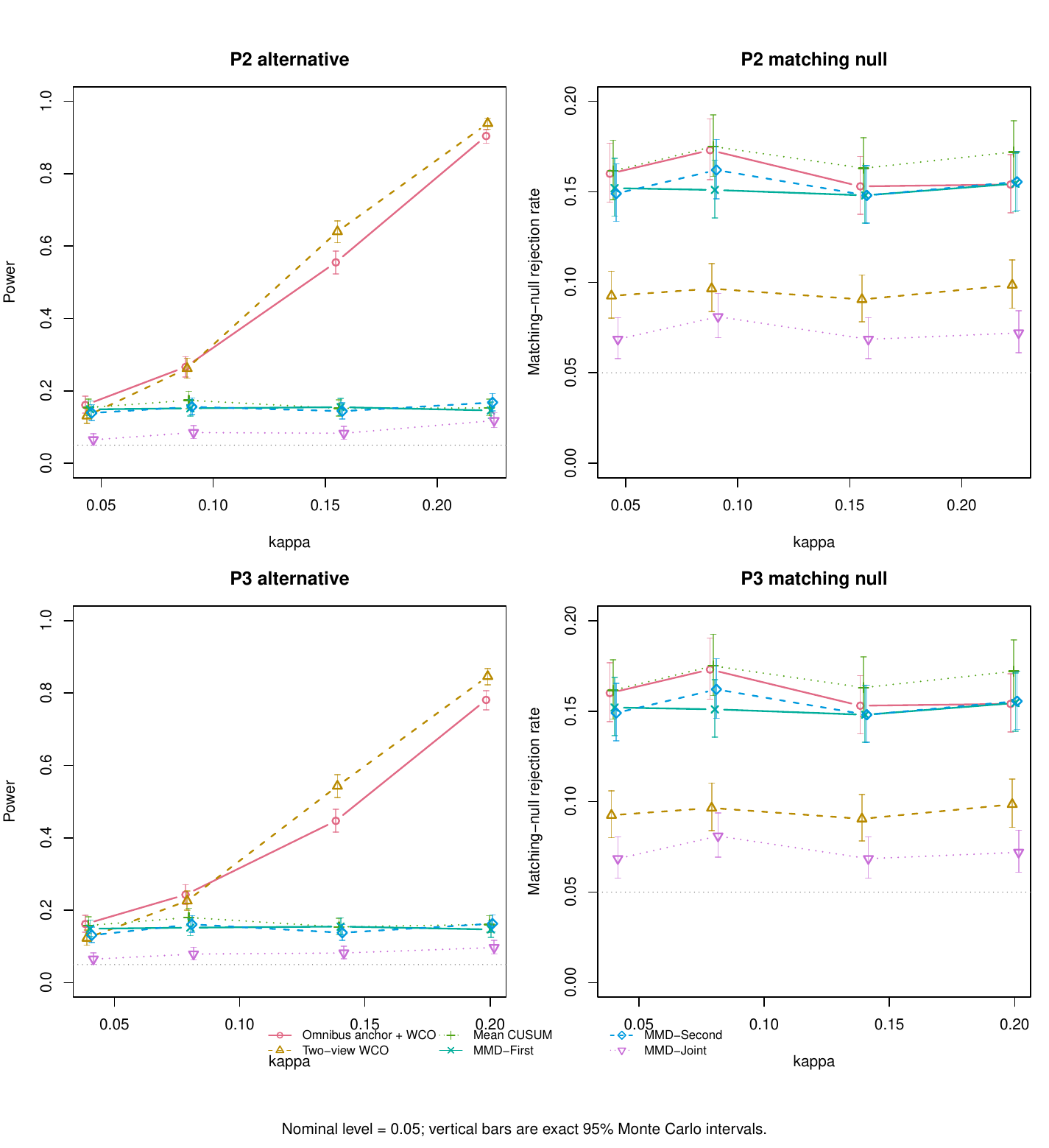}
\caption{
Empirical rejection probabilities under the targeted P2 and P3 alternatives
of Simulation I, plotted against the population WCO signal $\kappa$.
The left panels show empirical power, and the right panels show the
corresponding rejection probabilities under matching null cells.
Vertical bars are exact 95\% binomial Monte Carlo intervals, and the
horizontal dotted line denotes the nominal level 0.05.
}
\label{fig:simulation_I_power}
\end{figure*}

Table~\ref{tab:simulation_I_summary} summarizes the empirical rejection
probabilities under the baseline setting, and
Figure~\ref{fig:simulation_I_power} displays the power curves for the
target-aligned P2 and P3 alternatives. Table~\ref{tab:simulation_I_localization}
further reports the detection and localization performance under the
strongest targeted alternatives.

\begin{table*}[ht]
\centering
\caption{
Simulation I results under the baseline setting
$n=500$, $d=5$, $\varphi=0.3$, and $\tau_0=0.5$.
Entries are empirical rejection probabilities.
P1 evaluates null calibration; P2 and P3 are targeted
cross-view covariance alternatives; P4 contains target-orthogonal
location and nonlinear changes.
}
\label{tab:simulation_I_summary}
\small
\setlength{\tabcolsep}{4.2pt}
\resizebox{\textwidth}{!}{
\begin{tabular}{lccccccccccc}
\toprule
& \multicolumn{1}{c}{P1}
& \multicolumn{4}{c}{P2: sparse cross-view change}
& \multicolumn{4}{c}{P3: low-rank cross-view change}
& \multicolumn{2}{c}{P4: negative controls}
\\
\cmidrule(lr){2-2}
\cmidrule(lr){3-6}
\cmidrule(lr){7-10}
\cmidrule(lr){11-12}
Method
& Null
& $\kappa=0.0447$
& $\kappa=0.0894$
& $\kappa=0.1565$
& $\kappa=0.2236$
& $\kappa=0.0400$
& $\kappa=0.0800$
& $\kappa=0.1400$
& $\kappa=0.2000$
& Location
& Nonlinear
\\
\midrule
Omnibus anchor + WCO
& 0.1530
& 0.1610 & 0.2660 & 0.5550 & 0.9040
& 0.1620 & 0.2430 & 0.4470 & 0.7810
& 1.0000 & 0.7040
\\
Two-view WCO
& 0.0905
& 0.1310 & 0.2620 & 0.6400 & 0.9390
& 0.1230 & 0.2260 & 0.5430 & 0.8460
& 0.0730 & 0.1030
\\
Mean CUSUM
& 0.1630
& 0.1540 & 0.1740 & 0.1520 & 0.1530
& 0.1580 & 0.1800 & 0.1540 & 0.1610
& 1.0000 & 0.1680
\\
MMD--First
& 0.1480
& 0.1490 & 0.1520 & 0.1550 & 0.1460
& 0.1490 & 0.1520 & 0.1550 & 0.1470
& 0.9870 & 0.1570
\\
MMD--Second
& 0.1480
& 0.1390 & 0.1560 & 0.1440 & 0.1680
& 0.1310 & 0.1610 & 0.1380 & 0.1630
& 1.0000 & 0.8260
\\
MMD--Joint
& 0.0685
& 0.0650 & 0.0850 & 0.0830 & 0.1180
& 0.0650 & 0.0790 & 0.0820 & 0.0970
& 0.6890 & 0.3100
\\
\bottomrule
\end{tabular}
}
\begin{minipage}{0.98\textwidth}
\footnotesize
\textit{Notes:}
P2 changes a sparse set of entries in the cross-view covariance operator,
whereas P3 introduces a dense rank-one perturbation with alternating signs.
P4-location changes the marginal means without changing the centered WCO
target. P4-nonlinear changes the second-view distribution through a centered
quadratic relationship while preserving zero linear cross-covariance.
The nominal level is 0.05. The elevated null rejection rates under
$\varphi=0.3$ indicate that the dependence-bandwidth calibration requires
further adjustment. Consequently, raw rejection probabilities should be
interpreted together with the matching-null results.
\end{minipage}
\end{table*}

\begin{table}[H]
\centering
\caption{
Detection and localization under the strongest targeted alternatives
in Simulation I. Conditional MAE is the mean value of
$|\widehat{\tau}-\tau_0|$ among rejected replications, and the final column
gives $\Pr(|\widehat{k}-k_0|\leq 0.05n)$.
}
\label{tab:simulation_I_localization}
\small
\setlength{\tabcolsep}{5pt}
\begin{tabular}{llrrrr}
\toprule
Scenario & Method
& $\kappa$
& Power
& Conditional MAE
& Within $0.05n$
\\
\midrule
P2
& Omnibus anchor + WCO
& 0.2236 & 0.9040 & 0.0638 & 0.4920
\\
P2
& Two-view WCO
& 0.2236 & 0.9390 & 0.0287 & 0.7930
\\[2pt]
P3
& Omnibus anchor + WCO
& 0.2000 & 0.7810 & 0.0670 & 0.4580
\\
P3
& Two-view WCO
& 0.2000 & 0.8460 & 0.0344 & 0.7370
\\
\bottomrule
\end{tabular}
\end{table}

The targeted P2 and P3 designs clearly distinguish Two-view WCO from
procedures based only on marginal means or marginal distributions. Under P2,
the rejection probability of Two-view WCO increases monotonically from
0.1310 at $\kappa=0.0447$ to 0.9390 at $\kappa=0.2236$. Under P3, it
increases from 0.1230 at $\kappa=0.0400$ to 0.8460 at
$\kappa=0.2000$. These increases are substantial relative to the WCO null
rejection probability of 0.0905. In contrast, Mean CUSUM, MMD--First, and
MMD--Second remain close to their corresponding null rejection probabilities
throughout the P2 and P3 signal ranges. MMD--Joint exhibits only a modest
increase under the strongest signals, reaching 0.1180 in P2 and 0.0970 in
P3. This pattern is expected because P2 and P3 preserve the marginal means
and marginal covariance matrices while changing the cross-view covariance
targeted by the WCO $U$-statistic.

The results illustrate the main advantage of the proposed WCO construction.
Rather than searching for an arbitrary marginal discrepancy, WCO aggregates
pairwise products of changes across the two views and therefore concentrates
its power on changes in the cross-view covariance operator. This targeted
construction is effective for both the sparse entrywise perturbation in P2
and the dense rank-one perturbation in P3. The increasing power curves also
show that the WCO statistic responds systematically to the population signal
$\kappa$, rather than merely exploiting an isolated simulation
configuration.

The asymmetric Omnibus anchor + WCO procedure also gains substantial power
under the targeted alternatives. Its rejection probability reaches 0.9040
in the strongest P2 cell and 0.7810 in the strongest P3 cell. These values
are lower than the corresponding WCO rejection probabilities of 0.9390 and
0.8460. Part of this difference is expected because the composite procedure
requires joint calibration over several components, whereas WCO alone
concentrates on the functional changed by P2 and P3.

A more consequential difference appears in localization. In the strongest
P2 cell, Two-view WCO has a conditional change-fraction MAE of 0.0287,
compared with 0.0638 for the composite procedure. The corresponding
probabilities of localization within $0.05n$ are 0.7930 and 0.4920. Under
P3, WCO has a conditional MAE of 0.0344 and a within-$0.05n$ probability of
0.7370, whereas the composite procedure has a conditional MAE of 0.0670 and
a within-$0.05n$ probability of 0.4580. Thus, when the true change is aligned
with the WCO target, the WCO scan provides both higher detection probability
and substantially more accurate localization.

The weaker localization of the composite procedure follows from its
deliberately asymmetric construction. Although its rejection can be driven
by a strong WCO component, its reported location is anchored by the most
significant standardized MMD component. In P2 and P3, the MMD components
have relatively weak population signals because the marginal distributions
are preserved. The composite procedure is therefore useful as a robust
general detector, but its MMD-based location need not be as accurate as the
WCO maximizer under a specifically cross-view alternative. This distinction
supports reporting the omnibus detection result and the WCO diagnostic,
including its own estimated location and leading operator direction,
separately.

The P4 designs clarify the scope and specificity of WCO. Under the location
alternative, Mean CUSUM and the marginal MMD procedures have rejection
probabilities close to one, whereas Two-view WCO rejects with probability
0.0730, below its P1 rejection probability of 0.0905. Under the centered
nonlinear alternative, MMD--Second rejects with probability 0.8260, while
WCO rejects with probability 0.1030. Relative to the WCO null rejection
probability, the latter increase is small. These findings confirm that the
default WCO statistic is a targeted detector of linear cross-view covariance
change, rather than an omnibus detector of arbitrary distributional change.
The low WCO rejection rates in P4 are therefore an expected specificity
property rather than a loss of power against the registered target.

The omnibus anchor complements this specificity. It rejects with probability
1.0000 under the location alternative and 0.7040 under the nonlinear
alternative, demonstrating that the composite procedure retains sensitivity
to changes outside the WCO target. Taken together, the P2--P4 results support
the intended division of labor: the omnibus component detects broad
distributional changes, while WCO determines whether the change involves the
relationship between direct co-occurrence and shared-context representations.

The calibration results nevertheless require caution. Under the baseline
dependence level $\varphi=0.3$, the empirical P1 rejection probabilities are
0.1530 for the composite procedure, 0.1630 for Mean CUSUM, and 0.1480 for
both marginal MMD scans. Two-view WCO is more stable than these procedures
but still over-rejects, with empirical size 0.0905. Only MMD--Joint is
comparatively close to the nominal level, with rejection probability 0.0685.
Accordingly, the raw power values cannot be interpreted as fully
size-adjusted comparisons. The strongest evidence in favor of WCO is instead
the combination of its pronounced signal-dependent power increase, the lack
of corresponding increases for the marginal competitors, and its superior
localization under P2 and P3. The final simulation analysis will therefore
use a more conservative method-specific dependence-bandwidth calibration and
will continue to report matching-null results alongside every power curve.


\section{Real Data Analysis}\label{sec:realdata}

We illustrate the proposed method using a historical text corpus from the
journal \textit{New Youth}, published between 1915 and 1921.
\textit{New Youth} played a central role in the Chinese New Culture Movement,
which marked a major intellectual and linguistic transformation in modern
Chinese history. One of the defining features of this period was the movement
away from classical written Chinese toward vernacular language. This
transition was closely associated with the May Fourth Movement of 1919,
which is widely regarded as an important cultural and linguistic turning
point.

Our objective is to determine whether the corpus contains a statistically
detectable structural change and, if so, to estimate its timing directly from
the evolving lexical network. Rather than representing each text only by a
word-frequency vector, we construct time-indexed word-co-occurrence networks
and examine both direct lexical associations and shared-context structure.
The analysis uses a common vocabulary consisting of the $m=200$ retained
words. For each time-indexed text $\mathcal T_t$, two retained words are
treated as co-occurring when they appear within a textual span of
$h=10$ token positions. Applying this rule produces a symmetric weighted
co-occurrence matrix $W_t$.

Let $D_t$ be the diagonal degree matrix associated with $W_t$. We form the
degree-normalized adjacency matrix
$
S_t
=
(D_t^\dagger)^{1/2}
W_t
(D_t^\dagger)^{1/2}.
$
The first-order representation records normalized direct word
co-occurrences 
$
G_t^{(1)}
=
\operatorname{vech}_{\mathrm{off}}(S_t).
$
The second-order representation records shared lexical contexts through
normalized two-step paths:
$
G_t^{(2)}
=
\operatorname{vech}_{\mathrm{off}}
\left\{
\operatorname{off}(S_t^2)
\right\}.
$
Thus, the first view describes which words occur together directly, whereas
the second view describes which words are connected through common lexical
neighbors.

The two high-dimensional graph representations are centered, projected, and
whitened separately using the fixed PCA maps described in
Section~\ref{sec:method}. We retain three coordinates from each view, so that
$d_X=d_Y=3$. The resulting paired observations are
$$
X_t
=
P_X\{G_t^{(1)}-\mu_X\}
\in\mathbb R^3
\qquad\text{and}\qquad
Y_t
=
P_Y\{G_t^{(2)}-\mu_Y\}
\in\mathbb R^3.
$$
The centers, projection directions, and whitening factors are held fixed
throughout the change-point scan and all bootstrap replications. The
Two-view WCO procedure consequently tests for a change in the relationship
between the three retained direct-co-occurrence coordinates and the three
retained shared-context coordinates, rather than for a change in a single
two-dimensional PCA trajectory.

\begin{figure}[htbp]
\centering
\safeincludegraphics[width=0.6\linewidth]{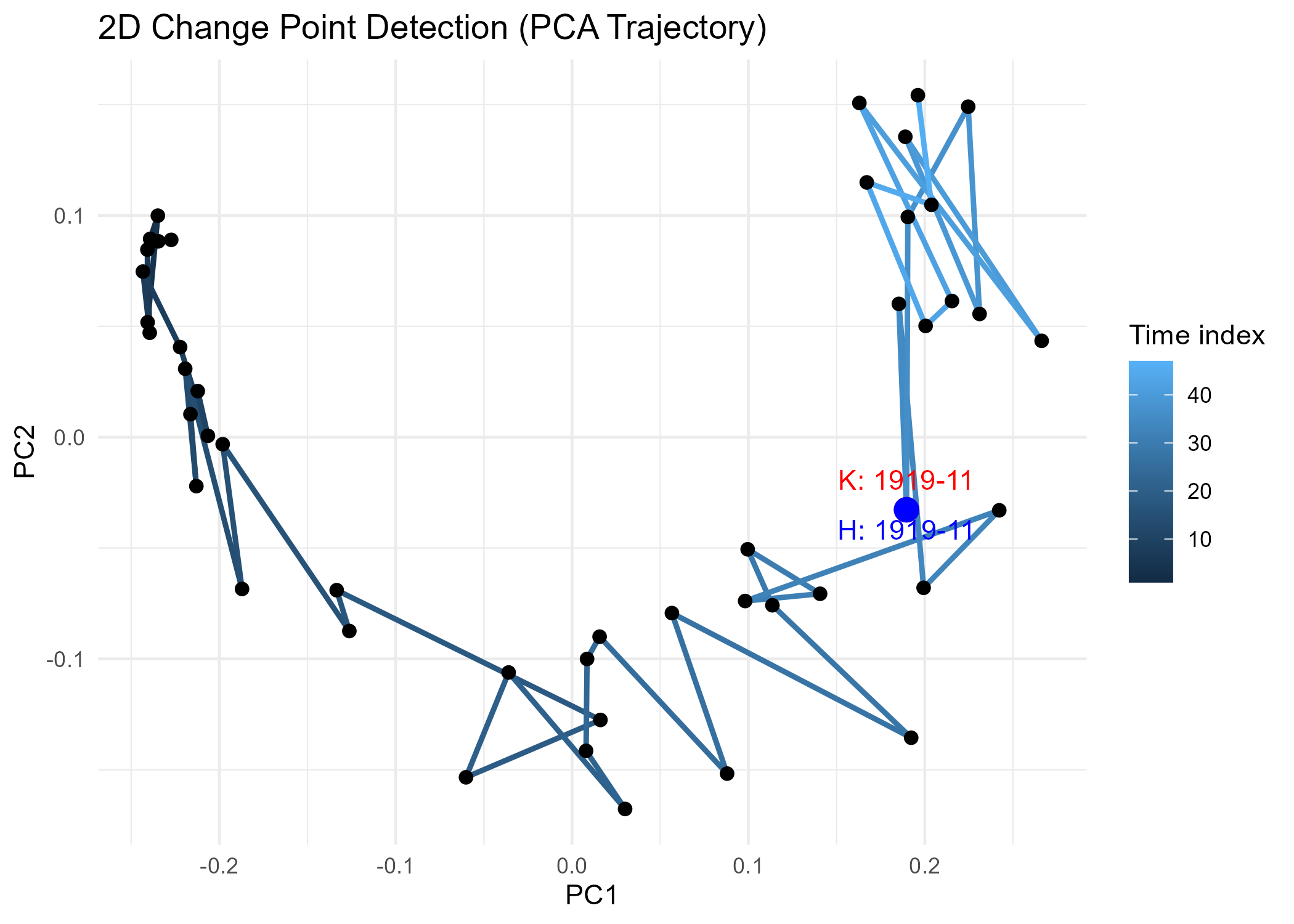}
\caption{Two-dimensional PCA representation of the \textit{New Youth} dataset. Each point corresponds to a document ordered in time, forming a trajectory that reflects linguistic evolution.}
\label{fig:trajectory_2d}
\end{figure}

The transformed observations $\{Z_t\}$ form a time-ordered sequence. As shown in Figure~\ref{fig:trajectory_2d}, the data exhibit a clear trajectory over time, suggesting a gradual but structured change in the underlying distribution.

\begin{figure}[htbp]
\centering
\safeincludegraphics[width=0.6\linewidth]{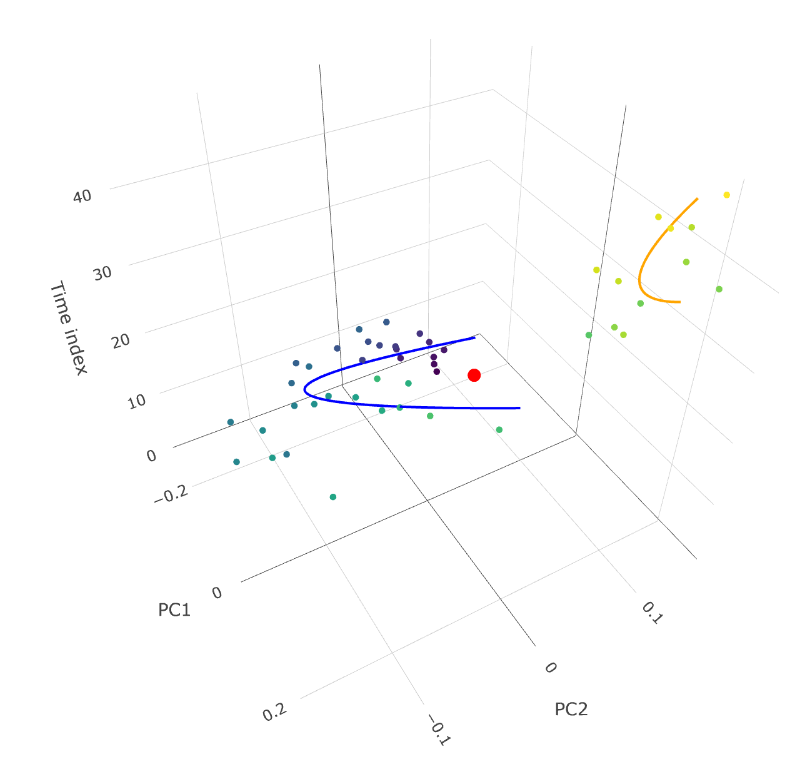}
\caption{Three-dimensional trajectory of the data with time as the third axis. A structural shift is visible in the later period.}
\label{fig:trajectory_3d}
\end{figure}

To further illustrate temporal dynamics, Figure~\ref{fig:trajectory_3d} presents a three-dimensional visualization with time as an additional axis. A noticeable shift in the trajectory can be observed in the later stage.

\begin{figure}[H]
\centering
\safeincludegraphics[width=0.6\linewidth]{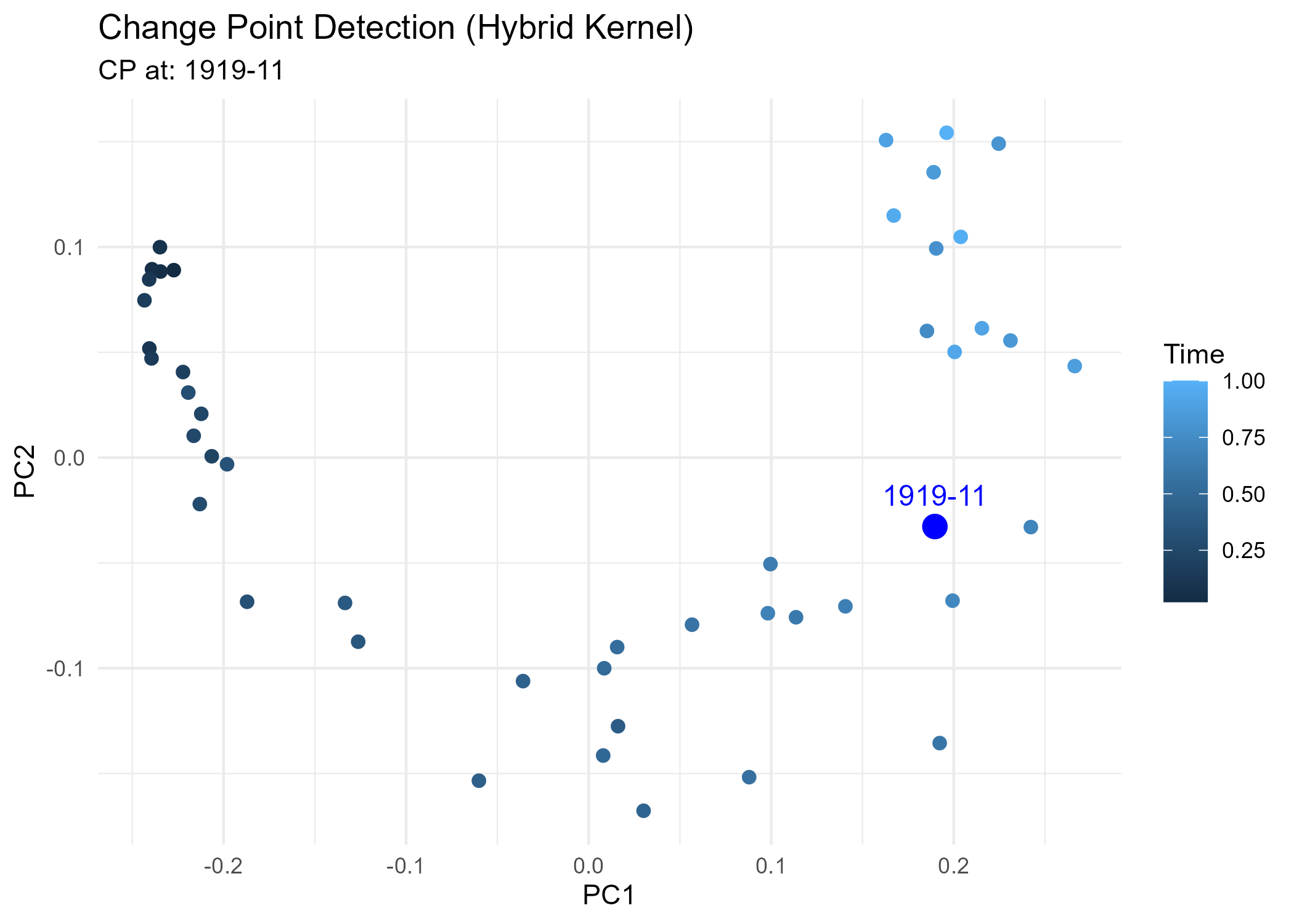}
\caption{Jointly calibrated scan for change-point detection. The peak corresponds to the estimated change-point.}
\label{fig:hybrid_cp}
\end{figure}

We apply the proposed jointly calibrated scan to the sequence $\{Z_t\}$ and
use its characteristic-kernel component as the localization anchor.  The
resulting combined statistic is shown in Figure~\ref{fig:hybrid_cp}, where a
clear peak is observed.

The detected change-point corresponds to November 1919.  A segment-centered 95\%
bootstrap stability interval is April 1919  and June 1920. Because the present theory establishes test calibration and localization but
not coverage of this interval construction, we report it descriptively and do
not label it a confidence interval.

The detected change-point aligns closely with the historical timeline of the May Fourth Movement. This period corresponds to a rapid shift toward vernacular Chinese, accompanied by substantial changes in writing style, syntax, and vocabulary. Notably, the detected change-point is obtained purely from the data without incorporating any historical information, yet it closely matches a well-known cultural turning point.
These results indicate that the proposed method captures both distributional and structural linguistic changes, demonstrating its applicability beyond controlled simulation settings.

The real data analysis demonstrates that:
(i) the proposed method effectively detects meaningful structural changes in high-dimensional data;
(ii) the estimated change-point is consistent with known historical events;
(iii) the jointly calibrated combination gives a clearer signal in this application.

\section{Conclusion}
\label{sec:conclude}

We have developed a representation-aware framework for offline change-point
detection in weakly dependent text networks. Each time-indexed text is
represented by a weighted word-co-occurrence network constructed on a common
vocabulary. From the degree-normalized adjacency matrix, we form two
complementary linguistic views: a first-order view describing normalized
direct word co-occurrences and a second-order view describing shared-context
relations through normalized two-step paths. Fixed projection and whitening
maps transform these representations into paired Euclidean coordinates.
The proposed weighted-concordance operator (WCO) $U$-statistic then measures
changes in the relationship between the two views. Under the primary
specification $\beta=\gamma=1$, its population target is twice the
cross-covariance between the direct-co-occurrence and shared-context
coordinates.

The targeted nature of WCO is central to its interpretation. WCO is not
intended to detect every possible distributional change. A marginal location
shift, a purely nonlinear change preserving linear cross-covariance, or
another alternative that leaves the WCO functional unchanged need not
produce a strong WCO signal. We therefore distinguish general detection from
mechanism-specific diagnosis. Characteristic Gaussian-kernel maximum mean
discrepancy procedures applied to the first-order, second-order, and
concatenated views provide omnibus evidence of distributional change, while
mean cumulative sum procedures provide benchmarks for location changes. In
the asymmetric composite procedure, the MMD components serve as the omnibus
anchor, whereas WCO determines whether the detected change involves the
relationship between direct lexical links and shared-context structure. The
leading singular directions of the segment-level WCO contrast further
identify the projected coordinates most responsible for that change.

A second contribution is to account explicitly for the fact that the network
representations are estimated from finite texts. Primitive conditions on
document length, token perturbations, vocabulary stability,
co-occurrence-window weights, graph mass, active node degrees, and fixed
pilot maps propagate text-recording error through the co-occurrence matrix,
the first- and second-order graph representations, the projected
coordinates, and the final WCO scan. This analysis provides an explicit
empirical-to-oracle representation rate instead of treating the constructed
network features as error-free observations. On this basis, we establish
oracle joint weak convergence of the WCO scan coordinates under weak temporal
dependence, validity of full-sample dependent multiplier calibration under
the no-change hypothesis, and asymptotic size control. Under a single
identifiable change and an explicit population separation condition, we
further establish detection consistency, consistency of the estimated change
fraction, and localization of the corresponding boundary. The separation
condition is essential because a mixed-segment WCO functional need not attain
its population maximum at the physical regime boundary.

The paired-view simulation study illustrates both the advantages and the
scope of the proposed statistic. Under sparse and dense low-rank changes in
cross-view covariance, Two-view WCO exhibits a pronounced increase in power
as the population WCO signal grows, while procedures based only on marginal
means or marginal distributions show little corresponding increase. Under
the strongest targeted alternatives, WCO also provides more accurate
localization than the omnibus composite procedure. Conversely, under
target-orthogonal location and centered nonlinear changes, WCO remains close
to its null rejection level, whereas the appropriate CUSUM or MMD procedures
retain high power. These findings support the intended division of labor
between omnibus detection and WCO-based structural diagnosis. They also show
that valid dependence calibration is critical: under the baseline dependent
design, several procedures over-reject, so power must be interpreted together
with matching-null rejection rates and method-specific bandwidth
calibration.

The historical application to the \emph{New Youth} corpus provides a
substantive illustration of the framework. Using a common vocabulary of
$m=200$ retained words, a co-occurrence span of $h=10$, and three projected
coordinates for each view, the analysis detects a linguistic change in
November 1919. The descriptive segment-centered bootstrap stability interval
extends from April 1919 to June 1920. This period is consistent with the
linguistic transformation surrounding the May Fourth and New Culture
movements. Beyond identifying a date, the two-view analysis distinguishes
changes in direct word co-occurrence from changes in shared-context structure
and assesses whether the relationship between these linguistic structures
changed.

Several extensions remain open. First, the present formal theory concerns a
single change. Recursive, interval-based, or penalized segmentation provides
a natural route to multiple-change analysis, but uniform recovery guarantees
for the resulting procedure require separate theoretical development.
Second, the vocabulary dimension and retained projection dimensions are
treated as fixed. A growing-vocabulary theory would require
dimension-dependent concentration, degree-stability conditions, and control
of the pilot projection maps. Third, oracle equivalence requires the
finite-document representation error to vanish sufficiently quickly relative
to the number of time periods. A noise-aware limit theory is needed when
finite-document error remains first-order rather than asymptotically
negligible. Fourth, the present default WCO targets linear cross-view
covariance. Developing and calibrating a richer family of nonlinear
concordance operators could extend diagnostic sensitivity while preserving a
clear population interpretation. Finally, extending the framework from
offline analysis to sequential monitoring, in the spirit of
\citet{Xie2013} and \citet{Leung2017}, is an important direction for future
work.

Overall, the proposed framework connects finite-text network construction,
two-view graph representations, dependent $U$-process inference, and
change-point analysis. Its principal contribution is not an omnibus statistic
that replaces all existing procedures, but a structured inferential strategy
that separates three questions: whether the represented text distribution
changed, where that change occurred, and whether it involved the relationship
between direct co-occurrence and shared-context structure. This separation
between omnibus detection, mechanism-specific diagnosis, and representation
uncertainty is potentially useful beyond text corpora whenever the
time-ordered objects supplied to a change-point procedure are themselves
estimated structured representations.

\section{Detailed Proof}
Detailed proof consists of proofs for every proposition, lemma, corollary, and theorem. And it is available upon request.




\bibliographystyle{plainnat}
\bibliography{ref}

\end{document}